\documentclass[11pt,a4paper]{article}

\usepackage[margin=0.8in]{geometry}
\usepackage{amsmath,amssymb,amsfonts,amsthm,mathtools}
\usepackage{thmtools}
\usepackage{physics}
\usepackage{braket}
\usepackage{bm}
\usepackage{bbm}
\usepackage{graphicx}
\usepackage{tcolorbox}
\usepackage{tabularx}
\usepackage{threeparttable}
\usepackage{float}
\usepackage{booktabs}
\usepackage{subcaption}
\usepackage{array}
\usepackage{multirow}
\usepackage{enumitem}
\usepackage{xcolor}
\usepackage{hyperref}
\usepackage[nameinlink,capitalize]{cleveref}
\usepackage[numbers,sort&compress]{natbib}
\usepackage{quantikz}
\usepackage{pdflscape}
\hypersetup{
  colorlinks=true,
  linkcolor=blue!60!black,
  citecolor=blue!60!black,
  urlcolor=blue!60!black
}

\usepackage{soul}
\usepackage{tikz}
\usetikzlibrary{shapes.geometric, arrows.meta, positioning}
\definecolor{green}{RGB}{0,150,0}
\definecolor{orange}{RGB}{255,150,79}
\definecolor{purple}{RGB}{209,79,255}

\declaretheorem[name=Theorem, numberwithin=section]{theorem}
\declaretheorem[name=Definition, sibling=theorem]{definition}
\declaretheorem[name=Proposition, sibling=theorem]{proposition}

\declaretheorem[name=Remark, sibling=theorem]{remark}

\newcommand{\C}{\mathbb{C}}
\newcommand{\R}{\mathbb{R}}

\newcommand{\diag}{\mathrm{diag}}

\newcommand{\be}{\begin{equation}}
\newcommand{\ee}{\end{equation}}
\newcommand{\I}{I}

\title{\bfseries From block-encoding to Generalized Quantum Signal Processing: Principles, Algorithms and Applications
}
\author{
Tal Gurfinkel$^{1}$, Kaushika De Silva$^{1,2}$, Anuradha Mahasinghe$^{1,2}$, Jens Renders$^{1,2}$, \\ Jack Blyth$^{1}$,
James Greenwell$^{1}$, Archie Butterworth$^{1}$,
Yusen Wu$^{1}$, Lyle Noakes$^{1}$, \\ Miloud Bessafi$^{2}$, Frederic Cadet$^{2}$, Jingbo Wang$^{1}$\footnote{corresponding author: jingbo.wang@uwa.edu.au}\\[0.6em]
\small $^{1}$Centre for Quantum Information, Simulation and Algorithms, \\
\small The University of Western Australia, Perth, Australia.\\
\small $^{2}$University Paris City and University of Reunion, France.\\
\small 
\small 
\small 
}
\date{\today}

\begin{document}
\maketitle

\begin{abstract}

Modern quantum algorithms are increasingly formulated as coherent procedures for implementing polynomial transformations of operators and singular values. This perspective provides a powerful and unifying language for quantum algorithm design, connecting a wide range of distinct problems through five closely related key tools: block-encoding, qubitization, QSP, QSVT and GQSP. Block-encoding embeds non-unitary matrices into larger unitaries; qubitization converts block-encodings into structured operators; QSP, QSVT and GQSP enable polynomial transformations with near-optimal query complexity. Together, these techniques form a general toolkit for transforming matrix functions into implementable quantum circuits. This paper develops these techniques from first principles as a unified framework for constructing quantum algorithms. We apply this framework to representative applications to highlight design principles and demonstrate how distinct algorithms can be constructed from a unified sequence of operator transformations.
A central contribution is a systematic decision workflow for selecting the appropriate approach according to the operator structure and the desired transformation polynomial. This perspective clarifies when direct GQSP or through qubitization, or Laurent expansion, or QSVT is most appropriate. We organize algorithmic design into an end-to-end pipeline: identifying the target matrix function, constructing an appropriate block-encoding, determining the relevant spectral domain, designing a polynomial or Laurent-polynomial approximation, synthesizing the phase factors, and translating the transformation into an executable quantum circuit. By applying this unified framework to example applications, we showcase a practical methodology for reasoning, designing, and implementing quantum algorithms based on polynomial transformations.

\end{abstract}

\newpage

\setcounter{tocdepth}{1} 
\tableofcontents

\section{Introduction}

Quantum computation is intrinsically unitary, whereas most computational tasks in scientific computing, optimization, statistics, and machine learning are naturally expressed in terms of non-unitary linear-algebraic operations.  Solving a linear system uses the action of $A^{-1}$; simulating diffusion involves $e^{-tA}$; evaluating Green's functions requires resolvents $(zI-A)^{-1}$; eigenvalue filtering uses approximate spectral projectors; and optimization algorithms often require polynomial, rational, or exponential functions of matrices. A central challenge in quantum algorithm design is therefore to implement useful transformations of matrices while respecting the unitary structure of quantum mechanics.

The modern approach to this challenge is built upon a small number of unifying primitives. First, a matrix $A$ is embedded into a larger unitary operator through a \emph{block-encoding}~\cite{gilyen2019quantum}. Second, the block-encoded operator is transformed into a structured quantum walk via \emph{qubitization}~\citep{low2017optimal, low2019hamiltonian}, whose eigenphases encode the spectral information of the underlying matrix, enabling efficient Hamiltonian simulation and spectral estimation. 
Third, polynomial transformations of this spectral data are implemented using \emph{quantum signal processing} (QSP) or its extension~\cite{low2017optimal, low2019hamiltonian}, \emph{quantum singular value transformation} (QSVT)~\cite{gilyen2019quantum, martyn2021grand}. More recently, QSP was generalized into an alternative framework, termed \emph{generalized quantum signal processing} (GQSP)~\citep{motlagh2024generalized, berry2024doubling}, which extends the signal-processing framework to broader classes of operators and polynomial transformations, thereby simplifying certain synthesis procedures and expanding the range of implementable matrix functions. This development has led to the two frameworks existing in parallel, while the differences between them are largely technical. Therefore, there is a need to clarify which framework ought to be used to solve which problems, and that is the primary goal of this paper.

The signal processing framework is powerful because it separates quantum algorithm design into two conceptually distinct components: (1) Data access and encoding: construct a block-encoding of the matrix or operator of interest, and (2) Functional transformation: approximate a desired function $f$ by a polynomial or related transformation and implement $f(A)$ through QSP/QSVT/GQSP. This separation mirrors classical numerical linear algebra, where the representation of a matrix and the approximation of a matrix function are distinct design choices. Furthermore, the signal processing approaches of QSVT and GQSP often achieve optimal or near-optimal complexity when appropriately implemented. The complexity of these approaches is largely dependent on the choice of block-encoding and the polynomial approximation used. This simplifies the determination of quantum circuit complexity when using the signal processing algorithms.

This paper focuses on the following foundational frameworks:
\begin{enumerate}[label=\arabic*.]
  \item \textbf{Block-encoding:} The representation of matrices as sub-blocks of unitaries.
  \item \textbf{Qubitization:} A method for converting a block-encoding into a quantum walk with eigenphases related to encoded eigenvalues. 
  \item \textbf{QSP:} Quantum signal processing; a sequence of phase rotations that implements bounded polynomial transformations of scalar signals. 
  \item \textbf{QSVT:} Quantum singular value transformation; a general framework for applying polynomial transformations to singular values of block-encoded matrices.
  \item \textbf{GQSP:} Generalized quantum signal processing; extensions of QSP that use more flexible single-qubit rotations or generalized polynomial structures.
\end{enumerate}
This paper focuses on the underlying mathematical theory, algorithmic design principles, and practical aspects of numerical implementation.

\subsection{Historical Context}

\subsubsection{From Hamiltonian simulation to polynomial methods}

Hamiltonian simulation asks for the implementation of
$e^{-iHt}$,
where $H$ is a Hermitian operator describing the energy of a quantum system.  It was one of the original motivations for quantum computation.  Early algorithms relied on product formulas, particularly Trotter--Suzuki decompositions.  Later methods introduced sparse-Hamiltonian oracles, linear combinations of unitaries, truncated Taylor series, and quantum walks \citep{lloyd1996universal,berry2007efficient,childs2012LCU, berry2015simulating}.

A major conceptual shift occurred when Hamiltonian simulation was reinterpreted as a polynomial approximation problem.  Instead of decomposing $e^{-iHt}$ directly, one can approximate the scalar function $e^{-itx}$ on the spectrum of $H$ and then implement the corresponding polynomial transformation.  QSP made this viewpoint precise and showed that optimal Hamiltonian simulation can be achieved through carefully chosen single-qubit rotations \citep{low2017optimal}.  Qubitization then supplied a particularly clean way to expose spectral information from a block-encoded Hamiltonian~\citep{low2019hamiltonian}.

\subsubsection{From QSP to QSVT}

Quantum Signal Processing (QSP) was introduced by Low and Chuang~\cite{low2017optimal}. Originally conceptualized as a highly efficient method for Hamiltonian simulation, QSP works by interleaving a series of single-qubit rotations with a signal operator. This allowed the application of precise polynomial transformations to the eigenvalues of a block-encoded scalar, essentially operating within a compact $2 \times 2$ SU(2) matrix space. By manipulating these quantum phases, QSP proved that complicated functions could be synthesized on a quantum computer with optimal gate complexity, laying the groundwork for a major shift in how quantum algorithms are structured.

Shortly after introducing Quantum Signal Processing (QSP), Low and Chuang published their influential paper ``Hamiltonian Simulation by Qubitization"~\cite{low2019hamiltonian}, establishing a powerful new framework for quantum linear algebra. While QSP provided a method for implementing polynomial transformations of single-qubit rotations, qubitization supplied the crucial mechanism for applying these transformations to general operators. Building on ideas from discrete-time quantum walks, qubitization transforms a block-encoded operator into a structured quantum walk whose invariant two-dimensional subspaces encode the spectral information of the original matrix. In particular, the eigenvalues of the encoded operator are mapped to eigenphases of the walk operator, converting spectral information into a form that can be manipulated coherently on a quantum computer.

The central insight of qubitization is that each spectral component of a high-dimensional operator can be represented within an effective (SU(2)) subspace, thereby reducing the problem to a collection of single-qubit rotations. This process effectively “qubitizes” the operator, allowing the polynomial-transformation machinery of QSP to act independently on each spectral component. As a result, block-encoding, qubitization, and QSP together form a remarkably general framework for implementing matrix functions, Hamiltonian simulation, and spectral transformations with near-optimal query complexity. Historically, qubitization can be viewed as a natural evolution of quantum-walk techniques, elevating quantum walks from problem-specific algorithmic tools to a universal framework for operator manipulation and quantum linear algebra.

Quantum Singular Value Transformation (QSVT) was developed a few years later by Gilyén et al.~\cite{gilyen2019quantum} as a sweeping generalization of the original QSP framework. While QSP is inherently restricted to scalars or single-qubit operations, QSVT elevates the underlying algebraic mechanics to handle large, arbitrary, and even non-square matrices. By combining QSP with qubitization, along with deep insights into the structure of matrices, their block-encodings and their singular values, Gilyén et al. were able to remove the unitarity restrictions present in QSP and shift the paradigm of signal processing into the singular value domain. 

By embedding a target matrix into a larger unitary operator via block-encoding, QSVT applies a polynomial transformation directly to the singular values of that matrix all at once, transforming the entire operator in a single, unified framework. Ultimately, QSP is the foundational operation of QSVT, which acts on each subspace spanned by a singular vector of the matrix. By lifting the scalar constraints of QSP into a high-dimensional matrix landscape, QSVT succeeded in unifying a vast library of seemingly disparate quantum algorithms under a single conceptual roof. Today, classic algorithms such as Grover’s search and quantum linear-system solvers can be understood as tailored instances of the broader QSVT framework, which grew out of the development of quantum signal processing (QSP) just a few years earlier.

Important precursors to these ideas emerged in prior research on the efficient quantum implementation of structured matrices. Mahasinghe and Wang~\cite{mahasinghe2016efficient} showed how non-unitary Toeplitz and Hankel operators can be embedded into larger circulant matrices and implemented through unitary dilation, closely anticipating the modern block-encoding framework. 
Zhou and Wang~\cite{zhou2017QFTC} developed a complementary Fourier-based construction in which a circulant operator is realized within a larger unitary circuit and recovered by projection onto an ancilla subspace. More explicitly, in a separate paper~\cite{zhou2017efficient}, Zhou and Wang expressed circulant matrices as linear combinations of unitary shift operators and constructed a PREPARE–SELECT–PREPARE† circuit in which the target matrix appears as an ancilla block of a larger unitary. Together, these works anticipated key elements of modern block-encoding, including unitary dilation, structural embedding, and linear combinations of unitaries.

\subsubsection{Recent generalizations}

Standard QSP and QSVT impose several technical constraints on the class of polynomial transformations that can be implemented. In particular, the target polynomial must satisfy specific parity requirements and boundedness conditions on the interval of interest. These constraints are not arbitrary limitations of the framework; rather, they arise fundamentally from the unitarity of quantum evolution and the SU(2) structure underlying the signal-processing sequence. The remarkable consequence is that any polynomial satisfying these conditions can be implemented with near-optimal query complexity. However, the same constraints can make algorithm design challenging, especially when constructing approximations to matrix functions that do not naturally satisfy the required symmetry. In practice, considerable effort is often devoted to designing suitable polynomial approximations and synthesizing the corresponding phase factors that realize them.

To overcome some of these difficulties, an extension of QSP by Motlagh and Wiebe~\citep{motlagh2024generalized}, known as generalized quantum signal processing (GQSP) has recently emerged. This approach broadens the standard QSP paradigm by introducing more general signal processing operators, generalized rotation sequences, Laurent-polynomial representations, and related transformation techniques. By enlarging the space of admissible transformations, GQSP can simplify polynomial design, reduce the complexity of phase synthesis, and provide more direct constructions for certain matrix functions. In some cases, functions that require intricate polynomial approximations within the standard QSP/QSVT framework can be implemented more naturally using generalized constructions. Recent developments have already demonstrated improved algorithms for Hamiltonian simulation, matrix inversion, and other matrix-function implementations, while retaining many of the efficiency guarantees that make QSP and QSVT so powerful \citep{berry2024doubling}. Although the theory is still developing rapidly, GQSP is increasingly viewed as a natural extension of the polynomial-transformation framework that underpins modern quantum linear algebra, offering additional flexibility without sacrificing the fundamental advantages of the original approach.

A recent development by Mahasinghe \emph{et al.}~\cite{mahasinghe2025hermitian} demonstrates that certain Hermitian matrix functions can be implemented without explicitly constructing a block-encoded input signal. Instead, the authors exploit the \textit{symmetric-unitary decomposition} $H=\frac{1}{2}(U+U^{\dagger})$ of Hermitian contraction $H$ to construct a block-encoding of the target output polynomial transformations of $H$ through a coherent combination of a pair of appropriately chosen GQSP sequences, with signal unitaries $U$ and $U^{\dagger}$ respectively. Compared with conventional QSVT-based approaches, the framework avoids explicit block-encoding of the input operator leading to reduced ancilla overhead and simplified circuit architecture. The approach extends to settings in which the input  operator admits a Laurent polynomial representation in a unitary $U$. Although these methods apply only to a restricted class of Hermitian operators where $U$ in the decomposition is efficiently implementable, it highlights a promising direction beyond the traditional block-encoding paradigm and illustrates the growing flexibility of GQSP-based matrix-function synthesis. 

\addtocontents{toc}{\string\setcounter{tocdepth}{2}}

\section{Quantum Linear Algebra Framework}

\subsection{Mathematical Preliminaries and Notation}

Let $A\in\C^{m\times n}$ be a matrix.  Its operator norm is denoted by $\|A\|$.  
Any matrix $A\in\C^{m\times n}$ has a singular value decomposition, which we denote by $A = W\Sigma V^\dagger$ or
\be
  A = \sum_j \sigma_j \ket{u_j}\bra{v_j},
\ee
where $\sigma_j\ge 0$ and $\{\ket{u_j}\}$ and $\{\ket{v_j}\}$ are the left and right singular vectors. 
If $A$ is Hermitian, its eigenvalue decomposition is
\be
  A = \sum_j \lambda_j \ket{\psi_j}\bra{\psi_j}.
\ee
A function $f$ applied to a Hermitian matrix is defined by functional calculus:
\be
  f(A)=\sum_j f(\lambda_j)\ket{\psi_j}\bra{\psi_j}.
\ee

The signal processing frameworks reduce many algorithmic tasks to polynomial approximation. Chebyshev polynomials are especially important in this application.  They are defined by
\be
  T_k(x)=\cos(k\arccos x), \qquad x\in[-1,1].
\ee
They obey the recurrence
\be
  T_{k+1}(x)=2xT_k(x)-T_{k-1}(x),
\ee
with $T_0(x)=1$ and $T_1(x)=x$.  Chebyshev expansions are numerically stable and often near-optimal for approximating smooth functions on intervals.

\subsection{Block-encoding}
\begin{definition}[Block-encoding]\label{def:block_encoding}
Let $A\in\C^{n\times n}$ be an arbitrary square matrix. We say that a unitary $U$ is an $(\alpha,a,\epsilon)$-block-encoding of $A$ if
\be
  \left\|A-\alpha(\bra{0^a}\otimes I)U(\ket{0^a}\otimes I)\right\|\le \epsilon.
\ee
If $\epsilon=0$, the block-encoding is exact.
\end{definition}

\begin{remark}
Rectangular matrices can be block-encoded by first being embedded in the top left corner of the zero matrix.
Zero padding is also useful to obtain dimensions that are powers of two, for implementations using qubits.
\end{remark}

Equivalently, up to error, $U$ has the block form
\be
  U \approx
  \begin{pmatrix}
    A/\alpha & * \\
    * & *
  \end{pmatrix}.
\ee
Block-encoding is a fundamental interface between arbitrary matrices and quantum circuits.  The normalization $\alpha$ is necessary because every sub-block of a unitary has norm at most one.  Therefore, $\alpha\ge \|A\|$ is required.

Given a state $\ket{\psi}$ prepared on the $\log_2 n$ input qubits, to successfully implement $A/\alpha$ in a quantum circuit using a block-encoding $U$, we must prepare the $a$ ancilla qubits in the $\ket{0^a}$ state, and then measure them in the $\ket{0^a}$ state after applying $U$. 
The probability of measuring the ancilla qubits as $\ket{0^a}$ is given by $\norm{A\ket{\psi}}^2/\alpha^2$. We highlight this fact, since it is fundamentally important to the application of signal processing techniques, where block-encodings are often used.
\begin{theorem}[Success Probability of Block-Encodings]\label{thm:success_probability}
    Given an $(\alpha,a,0)$-block-encoding $U$ of $A$ and a state $\ket{0^a}\ket{\psi}$, the probability of successfully implementing $A\ket{\psi}/\alpha$ is given by 
    \begin{equation}
        \frac{\norm{A\ket{\psi}}^2}{\alpha^2},
    \end{equation}
where the vector $l_2$-norm is defined by $\|A|\psi\rangle\|=\sqrt{\langle\psi|A^{\dagger}A|\psi\rangle}$. Furthermore, if $U$ is a $(\alpha,a,\epsilon)$-block-encoding of $A$ then the probability of successfully implementing $A\ket{\psi}/\alpha$ is bounded by 
    \begin{equation}
        \frac{\max\{0,\norm{A\ket{\psi}}-\epsilon\}^2}{\alpha^2}\leq P_\text{success} \leq \frac{(\norm{A\ket{\psi}}+\epsilon)^2}{\alpha^2}.
    \end{equation}
\end{theorem}

Therefore, the success probability of a block-encoding scales on the order of $O(1/\alpha^{2})$. This should be taken into careful consideration when applying a block-encoding.

\begin{remark}
A block-encoding should be judged not only by its normalization $\alpha$, but also by its ancilla cost, gate complexity, query complexity, and compatibility with controlled operations and reflections about the ancilla state.
\end{remark}

\subsubsection{Constructing block-encodings}

There are several standard routes to block-encodings. We outline some approaches here, although every application will generally require careful consideration of the problem when block-encoding. Hence, the following block-encoding schemes should be taken as foundational building blocks, rather than state-of-the-art plug-and-play subroutines.

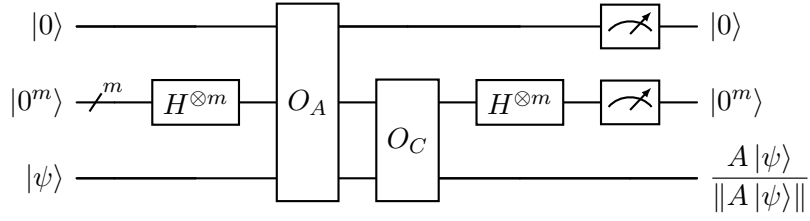
\begin{figure}
    \centering
\begin{quantikz}[
    row sep=0.4cm,
    column sep=0.5cm
]
\lstick{$\ket{0}$}
    & \qw
    & \qw
    & \gate[3]{O_A}
    & \qw
    & \qw
    & \meter{}
    & \rstick{$\ket{0}$}
\\
\lstick{$\ket{0^m}$}
    & \qwbundle{m}
    & \gate{H^{\otimes m}}
    &
    & \gate[2]{O_C}
    & \gate{H^{\otimes m}}
    & \meter{}
    & \rstick{$\ket{0^m}$}
\\
\lstick{$\ket{\psi}$}
    & \qw
    & \qw
    &
    &
    & \qw
    & \qw
    & \rstick{$\displaystyle
      \frac{A\ket{\psi}}{\norm{A\ket{\psi}}}$}\qw
\end{quantikz}
    \caption{Sparse-access block-encoding circuit for an $s$-sparse matrix, where $m=\lceil\log_2 s\rceil$. The upper two registers form the signal
register, and compression onto $\ket{0}\ket{0^m}$ yields the unnormalized post-selected branch corresponding to $A\ket{\psi}$, with
normalization factor $2^m$.
}
    \label{fig:sparse_access_block_encoding}
\end{figure}

\paragraph{Sparse-access block-encodings}
Sparse-access models are common in Hamiltonian simulation and linear-system algorithms. Suppose a matrix $A$ is $s$-sparse and accessible through oracles $O_C$ and $O_A$ that return the locations and values of nonzero entries respectively. Then one can construct a block-encoding with normalization scaling with sparsity and entry bounds. See Lemma 48 in~\cite{gilyen2019quantum} and Theorem 4.1 in \cite{camps2024blockencodings} for an explicit construction of such block-encodings. In this specific instance, they require $\log s$ additional ancilla, a single call to each of $O_A$ and $O_C$ as well as $2\log s$ single qubit gates (see \cref{fig:sparse_access_block_encoding}). Provided $O_A$ and $O_C$ can be implemented efficiently, this leads to an efficient block-encoding for sparse matrices~\cite{zhou2017efficient}. Camps et al. discuss several scenarios where these oracles can be implemented efficiently in \cite{camps2024blockencodings}.

\paragraph{Linear combination of unitaries}

\begin{figure}[ht]
    \centering
\begin{quantikz}[row sep=0.55cm, column sep=0.35cm]
        \lstick{$\ket{0^a}$}
        & \qwbundle{a}
        & \gate{\mathrm{PREP}}
        & \gate[wires=2]{\mathrm{SELECT}}
        & \gate{\mathrm{PREP}^{\dagger}}
        & \meter{}
        &\rstick{$\ket{0^a}$}
        \\
        \lstick{$\ket{\psi}$}
        & \qwbundle{m}
        & \qw
        &
        & \qw
        & \rstick{$\displaystyle
          \frac{A\ket{\psi}}{\norm{A\ket{\psi}}}$}\qw
    \end{quantikz}
    \caption{
LCU block-encoding circuit, where
$a=\lceil\log_2 L\rceil$. The upper register is the signal register,
and compression onto $\ket{0^a}$ yields the unnormalized post-selected
branch corresponding to $A\ket{\psi}$, with normalization factor
$\alpha$.} \label{fig:lcu_block_encoding}
\end{figure}
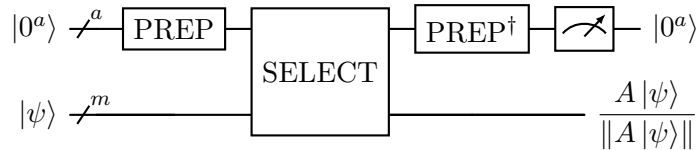

Suppose that your matrix can be written as a linear combination of unitaries. Namely,
\begin{equation}
    A = \sum_{j=0}^{L-1} \alpha_j U_j,
\end{equation}
where $U_j$ are efficiently implementable unitaries and $\alpha_j\ge0$.  Let
\begin{equation}
  \alpha=\sum_j \alpha_j.
\end{equation}
Define a state-preparation unitary $\mathrm{PREP}$ such that
\begin{equation}
  \mathrm{PREP}\ket{0}=\sum_j \sqrt{\alpha_j/\alpha}\ket{j},
\end{equation}
and a selection unitary
\begin{equation}
  \mathrm{SELECT}=\sum_j \ket{j}\bra{j}\otimes U_j.
\end{equation}
Then
\begin{equation}
  U_A=(\mathrm{PREP}^{\dagger}\otimes I)\mathrm{SELECT}(\mathrm{PREP}\otimes I),
\end{equation}
is an exact $(\alpha,\lceil\log_2L\rceil,0)$-block-encoding of $A$. This construction is based on the ``linear combination of unitaries (LCU)" method~\citep{childs2012LCU}, as shown in \cref{fig:lcu_block_encoding}. We also note that this implementation can be quite costly in general applications, due to the need to implement multi-controlled unitaries for the SELECT operator.

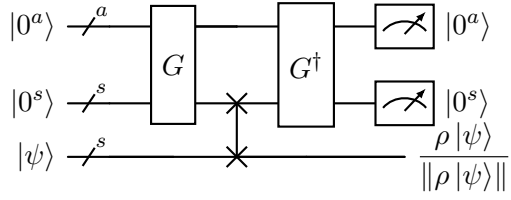
\begin{figure}[h]
    \centering
    \begin{quantikz}[
        row sep=0.4cm,
        column sep=0.55cm
    ]
    \lstick{$\ket{0^a}$}
        & \qwbundle{a}
        & \gate[2]{G}
        & \qw
        & \gate[2]{G^\dagger}
        & \meter{}\rstick{$\ket{0^a}$}
    \\
    \lstick{$\ket{0^s}$}
        & \qwbundle{s}
        &
        & \swap{1}
        &
        & \meter{}\rstick{$\ket{0^s}$}
    \\
    \lstick{$\ket{\psi}$}
        & \qwbundle{s}
        & \qw
        & \targX{}
        & \qw
        & \rstick{$\displaystyle
          \frac{\rho\ket{\psi}}
               {\norm{\rho\ket{\psi}}}$}\qw
    \end{quantikz}
    \caption{Density operator block-encoding via purification preparation. The
upper two registers form the signal register, and compression onto
$\ket{0^{a+s}}$ yields the unnormalized post selected branch $\rho\ket{\psi}$.}
    \label{fig:density_purification_block_encoding}
\end{figure}

\paragraph{Density operator block-encodings}
\label{sec:density_operators}
A block-encoding of a density matrix $\rho$ can be constructed from access to a purification preparation circuit of $\rho$. If a quantum circuit that prepares a purification of $\rho$ is available, meaning a unitary $G$ acting on registers $A$ and $S$ and such that $\operatorname{Tr}_A (G\ket{0}_A\ket{0}_S\bra{0}_S\bra{0}_A G^\dagger) = \rho$, then lemma 7 of \cite{low2019hamiltonian} states that a block-encoding of $\rho$ is obtained by
\begin{equation}
    U_\rho = (G^\dagger\otimes I_{S})(I_A \otimes \text{SWAP}_{S,S'})(G \otimes I_{S}).
\end{equation}
See \cref{fig:density_purification_block_encoding} for the corresponding quantum circuit.

\begin{figure}[h]
    \centering
\begin{quantikz}[
    row sep=0.4cm,
    column sep=0.55cm
]
\lstick{$\rho$}
    & \qwbundle{s}
    & \gate[2]{e^{-i\,\mathrm{SWAP}_s\,\Delta t}}
    & \rstick{\text{discard}}
\\
\lstick{$\sigma_k$}
    & \qwbundle{s}
    &
    & \rstick{$\sigma_{k+1}$}
\end{quantikz}
    \caption{One iteration of density matrix exponentiation. $\sigma_0 = \sigma$ and $\sigma_k \approx e^{-i\rho k \Delta t}\sigma e^{i\rho k\Delta t}$ for sufficiently small $\Delta t$. }
    \label{fig:density_exponentiation}
\end{figure}
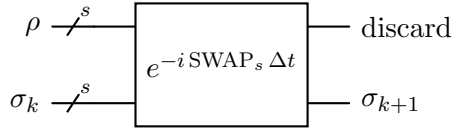

Without access to, or knowledge of a purification preparation operator $G$, an approximate block-encoding of $\rho$ can still be obtained, as long as multiple copies of state $\rho$ can be obtained. Density matrix exponentiation allows the implementation of a $(1, 0, \epsilon_0)$-block-encoding of the unitary $U_{\rho t} = e^{-i\rho t}$ by repeated partial swaps with the operand. A quantum circuit implementing a single ``exponential swap'' operator $e^{-i{\rm SWAP}_s\Delta t}$ followed by partial trace is shown in \cref{fig:density_exponentiation}. Each application of $U_{\rho t}$ requires $O\left(\frac{t^2}{\epsilon_0}\right)$ partial swaps \citep{lloyd2014quantum}. A block-encoding of $\rho$ itself can be obtained by transforming $U_{\rho t}$ using QSVT or GQSP, as shown in \cref{sec:practical_guide}.

\subsubsection{Basic algebra of block-encodings}

Block-encodings are composable.  If $U_A$ block-encodes $A$ and $U_B$ block-encodes $B$, then products and sums of $A$ and $B$ can be block-encoded using a combination of $U_A$ and $U_B$.

\begin{figure}[h]
    \centering
    \begin{quantikz}[
        row sep=0.17cm,
        column sep=0.35cm
    ]
    \lstick{$\ket{0}$}
        & \gate{H}
        & \gate[wires=3]{\scriptscriptstyle\mathrm{SELECT}_{U_A,U_B}}
        & \gate{H}
        & \meter{}\rstick{$\ket{0}$}
    \\
    \lstick{$\ket{0^a}_{\mathrm{anc}}$}
        & \qw
        &
        & \qw
        & \meter{}\rstick{$\ket{0^a}$}
    \\
    \lstick{$\ket{\psi}_{\mathrm{sys}}$}
        & \qw
        &
        & \qw
        & \rstick{$\displaystyle
          \frac{(A+B)\ket{\psi}}
               {\norm{(A+B)\ket{\psi}}}$}\qw
    \end{quantikz}

   \caption{Circuit for block-encoding $A+B$, given an
$(\alpha,a,\varepsilon_A)$ block-encoding of $A$ and a
$(\alpha,a,\varepsilon_B)$ block-encoding of $B$. The upper two registers form the signal register, and
compression onto $\ket{0}\ket{0^a}$ yields the unnormalized
post-selected branch corresponding to $(A+B)\ket{\psi}$, with
normalization factor $2\alpha$ and block-encoding error at most
$\varepsilon_A+\varepsilon_B$.
}
    \label{fig:lcu_sum_block_encoding}
\end{figure}
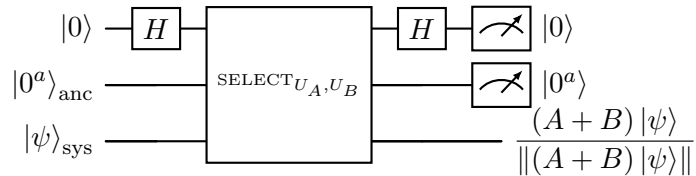

\begin{proposition}[Sum of block-encodings]\label{thm:block_encoding_addition}
    Let $U_A$ be a $(\alpha,a,\epsilon_A)$-block-encoding of an $s$-qubit matrix $A$ and $U_B$ be a $(\alpha,a,\epsilon_B)$-block-encoding of an $s$-qubit matrix $B$. Then 
    \begin{equation}
        U_{A+B} = (H\otimes I_{a+s})(\ketbra{0}{0}\otimes U_A + \ketbra{1}{1}\otimes U_B)(H\otimes I_{a+s}),
    \end{equation}
    is a $(2\alpha,a+1,\epsilon_A + \epsilon_B)$-block-encoding of $A+B$, as shown in \cref{fig:lcu_sum_block_encoding}.
\end{proposition}

More generally, linear combinations of block-encoded matrices can also be implemented similarly, by replacing the Hadamard gates with appropriate state preparation pairs and modifying the controlled unitary sandwiched between the two (see lemma 52 in \cite{gilyen2019quantum}).

\begin{figure}[h]
    \centering
\begin{quantikz}[row sep=0.45cm, column sep=0.35cm]
\lstick{$\ket{0^a}_A$}
    & \qw
    & \gate[wires=2]{\scriptscriptstyle U_A}
    & \meter{}
    & \rstick{$\scriptstyle\ket{0^a}$}
\\
\lstick{$\ket{\psi}_{\mathrm{sys}}$}
    & \gate[wires=2]{\scriptscriptstyle U_B}
    &
    & \qw
    & \rstick{$\scriptstyle
      AB\ket{\psi}\big/\norm{AB\ket{\psi}}$}\qw
\\
\lstick{$\ket{0^b}_B$}
    &
    & \qw
    & \meter{}
    & \rstick{$\scriptstyle\ket{0^b}$}
\end{quantikz}

    \caption{
    Circuit for block-encoding $AB$, given an
    $(\alpha,a,\varepsilon_A)$ block-encoding $U_A$ of $A$ and a
    $(\beta,b,\varepsilon_B)$ block-encoding $U_B$ of $B$. The upper and lower
    registers form the signal register, and compression onto
    $\ket{0^a}\ket{0^b}$ yields the unnormalized post-selected branch
    corresponding to $AB\ket{\psi}$, with normalization factor
    $\alpha\beta$ and block-encoding error at most
    $\alpha\varepsilon_B+\beta\varepsilon_A$.
    }
    \label{fig:block_encoding_product}
\end{figure}
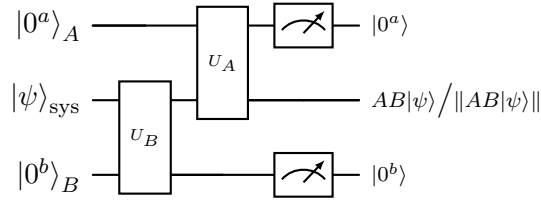

\begin{proposition}[Product of block-encodings \cite{gilyen2019quantum}]
If $U_A$ is an $(\alpha,a,\epsilon_A)$-block-encoding of $A$ and $U_B$ is a $(\beta,b,\epsilon_B)$-block-encoding of $B$, then  $(I_b\otimes U_A)(I_a\otimes U_B)$ is an $(\alpha\beta,a+b,\alpha\epsilon_B+\beta\epsilon_A)$ block-encoding of $AB$, as shown in \cref{fig:block_encoding_product}.
\end{proposition}

This proposition, combined with LCU or linear combinations of block-encodings allows us to construct polynomial transformations of a matrix $A$. That is, powers $A^k$ can be represented through repeated block-encodings. However, this approach creates significant overhead. The signal processing techniques, like GQSP are able to implement these block-encodings of polynomial transformations without such overhead.

\subsection{Qubitization}\label{sec:qubitization}

\subsubsection{Motivation}

Block-encoding provides unitary access to a generally non-unitary
operator by realizing \(A/\alpha\) as the compression of a larger
unitary \(U_A\) to designated signal subspaces. In general, however,
this construction does not impose a useful relationship between the
spectrum of \(U_A\) and the spectral data of \(A/\alpha\). For example,
the Hermitian Halmos dilation of a Hermitian contraction is an
involution and therefore has spectrum contained in \(\{-1,1\}\),
irrespective of the spectrum of the encoded operator. Thus, although
\(A/\alpha\) is faithfully represented in a designated block of
\(U_A\), its spectral information need not be directly accessible
through the eigenphases of \(U_A\).

Introduced by Low and Chuang~\cite{low2019hamiltonian}, qubitization
provides this additional spectral structure, primarily for Hermitian
signal operators, with an extension to normal operators through a
related construction. For a Hermitian signal operator, the precise
qubitization procedure depends on whether the block-encoding unitary
\(U_A\) is itself Hermitian. If it is, then \(U_A\), being both
Hermitian and unitary, is a reflection. Its product with a Grover-type
reflection about the signal subspace produces a \textit{qubitized walk operator} whose
dynamics decomposes into invariant subspaces of dimension at most two.
On each nontrivial subspace, the qubitized walk operator acts as a rotation whose
angle encodes the corresponding eigenvalue of \(A/\alpha\) \cite[Corollary~9]{low2019hamiltonian}. If \(A\) is Hermitian but \(U_A\) is not, Low and Chuang show that an
enlarged Hermitian block-encoding unitary can be constructed by
introducing an additional ancilla qubit and combining controlled
applications of \(U_A\) and \(U_A^\dagger\) \cite[Lemma~10]{low2019hamiltonian}. This reduces the qubitization
procedure to the preceding Hermitian-unitary setting. Their framework is further extended
to normal operators through a construction based on the polar
decomposition.

The discussion below is organized around these distinctions. First, the reflection-based construction for a Hermitian matrix with a Hermitian block-encoding is reviewed following \cite{low2019hamiltonian}. We then present the more general singular-value formulation of qubitization
introduced by Gilyén et al.~\cite{gilyen2019quantum}, which applies to
possibly non-Hermitian or rectangular matrices. In particular,
this formulation encompasses the qubitization of Hermitian matrices whose
block-encoding unitaries are not themselves Hermitian. In this setting, the single invariant subspace of the Hermitian construction is replaced by paired two-dimensional transition subspaces associated with the
corresponding right and left singular vectors. The block encoding unitary \(U_A\)
and its adjoint \(U_A^\dagger\) map between these subspaces, while the
associated singular value determines the rotation angle of the
resulting two-reflection walk.

\subsubsection{Qubitizing a Hermitian block-encoding of a Hermitian matrix}\label{sec:qubitising_hermitian_block_encoding}
\begin{definition}[Hermitian block-encoding]
A block-encoding $U_A$ of $A$ is said to be a Hermitian block-encoding if $U_A$ is Hermitian. \end{definition}
In this \lcnamecref{sec:qubitising_hermitian_block_encoding} we consider a $(\alpha,a,0)$ Hermitian block-encoding of a Hermitian matrix $H$ and denote the signal projector and its associated reflection by,
\begin{equation}
    \Pi =\ket{0^a}\bra{0^a}\otimes\I,
    \qquad R=2\Pi-\I.
\end{equation}
We define the corresponding \textit{qubitized walk operator} by
\begin{equation}\label{eq:qubitization_walk_operator}
    W_H=RU_H.
\end{equation}
Since $U_H$ is both Hermitian and unitary, $U_H^2=I$. Thus, $U_H$ and $R$ are reflections, and their product exhibits the two-dimensional rotation structure derived below.

Let $(\widehat{\lambda}_j,\ket{\psi_j})$ be an eigenpair of $H$, and
define
\begin{equation}
    \lambda_j
    =
    \frac{\widehat{\lambda}_j}{\alpha}
    \in[-1,1],
    \qquad
    \ket{v_j}
    =
    \ket{0^a}\ket{\psi_j},
\end{equation}
where $\alpha$ is a normalized factor enabling $\|H/\alpha\|\leq 1$.
The block-encoding relation implies
\begin{equation}
    \Pi U_H\ket{v_j}
    =
    \lambda_j\ket{v_j}.
\end{equation}
Thus, by probability preservation the component of
$U_H\ket{v_j}$ that is orthogonal to the signal subspace
$\operatorname{Ran}(\Pi)$ has norm $\sqrt{1-\lambda_j^2}$. For $\lvert\lambda_j\rvert<1$, 
\begin{equation}\label{eq:action_hermitian_block_encoding1}
    U_H\ket{v_j}
    =
    \lambda_j\ket{v_j}
    +
    \sqrt{1-\lambda_j^2}\ket{v_j^\perp},
\end{equation}
where $\ket{v_j^\perp}$ is normalized and orthogonal to the signal subspace. Using $U_H^2=\I$, one obtains
\begin{equation}\label{eq:action_hermitian_block_encoding2}
    U_H\ket{v_j^\perp}
    =
    \sqrt{1-\lambda_j^2}\ket{v_j}
    -
    \lambda_j\ket{v_j^\perp}.
\end{equation}
Hence the subspace
\begin{equation}
    \mathcal{V}_j
    =
    \operatorname{span}
    \left\{
        \ket{v_j},
        \ket{v_j^\perp}
    \right\},
\end{equation}
is invariant under both $U_H$ and $R$, and therefore under $W_H$. In the ordered basis $\{\ket{v_j},\ket{v_j^\perp}\}$, the two reflections have the matrix representations
\begin{equation}
    [U_H]_{\mathcal{V}_j}
    =
    \begin{pmatrix}
        \lambda_j & \sqrt{1-\lambda_j^2}\\
        \sqrt{1-\lambda_j^2} & -\lambda_j
    \end{pmatrix},
    \qquad
    [R]_{\mathcal{V}_j}
    =
    \begin{pmatrix}
        1 & 0\\
        0 & -1
    \end{pmatrix}.
\end{equation}
It follows that
\begin{equation}\label{eq:qubitization_as_rotation}
    [W_H]_{\mathcal{V}_j}
    =
    [R]_{\mathcal{V}_j}[U_H]_{\mathcal{V}_j}
    =
    \begin{pmatrix}
        \lambda_j & \sqrt{1-\lambda_j^2}\\
        -\sqrt{1-\lambda_j^2} & \lambda_j
    \end{pmatrix}
    =
    \begin{pmatrix}
        \cos\theta_j & \sin\theta_j\\
        -\sin\theta_j & \cos\theta_j
    \end{pmatrix},
\end{equation}
where $\theta_j=\arccos(\lambda_j).$ Thus, $W_H$ acts as a rotation on each invariant subspace $\mathcal{V}_j$. Its eigenvectors and corresponding eigenvalues are
\begin{equation}
    \ket{\varphi_j^\pm}
    =
    \frac{1}{\sqrt{2}}
    \left(
        \ket{v_j}
        \mp i\ket{v_j^\perp}
    \right),
    \qquad
    W_H\ket{\varphi_j^\pm}
    =
    e^{\mp i\theta_j}\ket{\varphi_j^\pm}.
\end{equation}

For the endpoint cases $\lambda_j=\pm1$, the orthogonal state invariant subspace reduces to $\operatorname{span}\{\ket{v_j}\}$, with
$W_H\ket{v_j}=\lambda_j\ket{v_j}$. 

\begin{theorem}[Qubitization principle for Hermitian block-encodings {\cite[Corollary~9]{low2019hamiltonian}}]
\label{thm:hermitian_qubitization}
Let $U_H$ be a $(\alpha,a,0)$ Hermitian block-encoding of a Hermitian operator
$H$ acting on $n$ qubits. For each eigenpair $(\widehat{\lambda}_j,\ket{\psi_j})$  of $H$ set
$  \lambda_j    =
    \frac{\widehat{\lambda}_j}{\alpha}
$. Define the qubitized walk operator
\begin{equation}
    W_H=(2\Pi-\I)U_H,
\end{equation}
where $\Pi=\ket{0^a}\bra{0^a}\otimes I_{2^n}$ is the projection on the signal subspace. Then for each $j$ with $|\lambda_j|<1$, the qubitized walk operator $W_H$ has an invariant subspace $\mathcal{V}_j$ of dimension two. On this subspace, its eigenvalues are
$e^{\pm i\theta_j}$, where $\cos\theta_j=\lambda_j$. Explicitly $\mathcal{V}_j=\mathrm{Span}\{\ket{v_j},\ket{v_j^{\perp}}\}$ where $\ket{v_j}=\ket{0^a}\ket{\psi_j}$, $\ket{v_j^{\perp}}=\frac{(I-\Pi)U_H\ket{v_j}}{\sqrt{1-\lambda^2_j}}$ and in this basis action of $W_H$ on $\mathcal{V}_j$ is represented by,
\begin{equation}
   [W_H]_{\mathcal{V}_j}= \begin{pmatrix}
        \cos\theta_j & \sin\theta_j\\
        -\sin\theta_j & \cos\theta_j
    \end{pmatrix}.
\end{equation}

For \(\lvert\lambda_j\rvert=1\), the corresponding invariant subspace
reduces to \(\operatorname{span}\{\ket{v_j}\}\), and
\begin{equation}
    W_H\ket{v_j}=\lambda_j\ket{v_j}.
\end{equation}
\end{theorem}

\subsubsection{Singular-value qubitization of general projected unitary encodings}
\label{sec:singular_value_qubitization}

\begin{definition}[Projected unitary encoding~\cite{gilyen2019quantum}]
\label{def:projected_unitary_encoding}
Let $A$ be an operator on Hilbert space $\mathcal{H}$, \(\Pi\) and \(\widetilde{\Pi}\) be orthogonal projectors, and $U_A$ be a unitary each acting on $\mathcal{H}$. Then
the triple \((U_A,\Pi,\widetilde{\Pi})\) is an exact projected unitary encoding with normalization factor $\alpha>0$ if

\begin{equation}
\label{eq:projected_unitary_encoding}
    \widetilde{\Pi}U_A\Pi
    =
    \frac{A}{\alpha}.
\end{equation}
\end{definition}

Projected-unitary encoding treats $A$ as an operator from $\operatorname{Ran}(\Pi)$ to
$\operatorname{Ran}(\widetilde{\Pi})$, accommodating different input and output dimensions. Define
the corresponding \textit{signal-space reflections} by
\begin{equation}
    R=2\Pi-\I,
    \qquad
    \widetilde{R}=2\widetilde{\Pi}-\I.
\end{equation}

Let $(\widehat{\sigma}_j,\ket{v_j},\ket{\widetilde{v}_j})$ be a
singular value triplet of $A$, so that
\begin{equation}\label{eq:singular_vector_pair}
    A\ket{v_j}
    =
    \widehat{\sigma}_j\ket{\widetilde{v}_j},
    \qquad
    A^\dagger\ket{\widetilde{v}_j}
    =
    \widehat{\sigma}_j\ket{v_j},
\end{equation}
with $ \ket{v_j}\in\operatorname{Ran}(\Pi)$ and $\ket{\widetilde{v}_j}\in
    \operatorname{Ran}(\widetilde{\Pi})$. Set
\begin{equation}
   \sigma_j
    =
    \frac{\widehat{\sigma}_j}{\alpha}
    \in[0,1].
\end{equation}
For $\sigma_j<1$, let $\ket{v_j^\perp}$ be the normalized component of $U_A\ket{v_j}$ that is orthogonal to $\operatorname{Ran}(\widetilde{\Pi})$, and $\ket{\widetilde{v}_j^\perp}$ be the normalized component of $U^\dagger_A\ket{\widetilde{v}_j}$ that is orthogonal to $\operatorname{Ran}({\Pi})$. The block-encoding relations then give,
\begin{equation}
    U_A\ket{v_j}=   \sigma_j\ket{\widetilde{v}_j}+\sqrt{1-\sigma_j^2}\ket{\widetilde{v}_j^\perp},
    \qquad
    U_A^\dagger\ket{\widetilde{v}_j}=\sigma_j\ket{v_j}+\sqrt{1-\sigma_j^2}\ket{v_j^\perp}.
    \label{eq:singular_value_encoding_actions}
\end{equation}

Accordingly, the pair of subspaces defined as
$\mathcal{V}_j^{\mathrm{in}}
    =
    \operatorname{span}
    \left\{
        \ket{v_j},
        \ket{v_j^\perp}
    \right\}$, $
    \mathcal{V}_j^{\mathrm{out}}
    =
    \operatorname{span}
    \left\{
        \ket{\widetilde{v}_j},
        \ket{\widetilde{v}_j^\perp}
    \right\}$, form a pair of \textit{transition spaces} for the actions of $U_A$ and $U_A^\dagger$. In particular, 
\begin{equation}\label{eq:singular_value_transition_maps}
    U_A\mathcal{V}_j^{\mathrm{in}}
    =
    \mathcal{V}_j^{\mathrm{out}},
    \qquad
    U_A^\dagger\mathcal{V}_j^{\mathrm{out}}
    =
    \mathcal{V}_j^{\mathrm{in}}.
\end{equation}
Thus, although neither transition space is generally invariant under
$U_A$, the pairing supports an invariant two-reflection qubitized walk operator defined by
\begin{equation}\label{eq:singular_value_qubiterates}
    W_A^{2\mathrm{ref}}
    =
    R U_A^\dagger\widetilde{R}U_A.
\end{equation}
Indeed, $ W_A^{2\mathrm{ref}}$ preserves
$\mathcal{V}_j^{\mathrm{in}}$ and in the  ordered bases $
    \left\{
        \ket{v_j},\ket{v_j^\perp}
    \right\}$ the matrix representation of the action of $W_A^{2\mathrm{ref}}$ on $\mathcal{V}_j$ is given by
\begin{equation}\label{eq:singular_value_qubitization_as_rotation}
    \left[W_A^{2\mathrm{ref}}\right]_{\mathcal{V}_j^{\mathrm{in}}}
      =
    \begin{pmatrix}
        \cos(2\theta_j) & \sin(2\theta_j)\\
        -\sin(2\theta_j) & \cos(2\theta_j)
    \end{pmatrix},
\end{equation}
where $\cos\theta_j=\sigma_j$. Therefore,
$e^{\pm 2i\theta_j}$, are eigenvalues of $W_A^{\mathrm{2ref}}$.
Hence, the eigenphases of the walk encode the scaled singular value
$\sigma_j$, while $U_A$ and $U_A^\dagger$ provide the transition
between the corresponding right and left singular-vector subspaces. 

In the special case where $A$ is Hermitian and the projected unitary encoding satisfies $\Pi=\widetilde{\Pi}$, the above construction can be formulated in terms of the eigenpairs of $A/\alpha$ rather than its singular triplets. To see this, for each eigenpair $(\lambda_j,\ket{v_j})$ of $A/\alpha$ with $|\lambda_j|<1$, set
\begin{equation}
    \ket{\widetilde{v}_j}=\ket{v_j},
    \qquad
    \ket{v_j^\perp}
    =
    \frac{(\I-\Pi)U_A^\dagger\ket{v_j}}
         {\sqrt{1-\lambda_j^2}},
    \qquad
    \ket{\widetilde{v}_j^\perp}
    =
    \frac{(\I-\Pi)U_A\ket{v_j}}
         {\sqrt{1-\lambda_j^2}}.
\end{equation}

Then \cref{eq:singular_value_encoding_actions} holds with $\sigma_j$ replaced by the eigenvalue $\lambda_j$. The remainder of the preceding analysis therefore carries through, and the resulting two-reflection walk operator $W_A^{2\mathrm{ref}}$ has the same rotation form as in
\cref{eq:singular_value_qubitization_as_rotation}, but now with
$\cos\theta_j=\lambda_j$ rather than $\cos\theta_j=\sigma_j$. The following theorem and subsequent remark summarize this construction. 

\begin{theorem}[Singular-value Qubitization principle~\cite{gilyen2019quantum}.]
\label{thm:singular_value_qubitization}
Let $(U_A,\Pi,\widetilde{\Pi})$ be an exact projected unitary encoding
of $A$ with associated normalization factor $\alpha>0$. For each singular value triplet $\left(\widehat{\sigma}_j,\ket{v_j},\ket{\widetilde{v_j}}\right)$ of $A$, set $\sigma_j=\frac{\widehat{\sigma}_j}{\alpha}\in[0,1]$
be the associated scaled singular value. Define the two-reflection qubitized walk operator
\begin{equation}
    W_A^{2\mathrm{ref}}
    =
    R U_A^\dagger\widetilde{R}U_A,
\end{equation}
where \begin{equation}
    R=2\Pi-\I,
    \qquad
    \widetilde{R}=2\widetilde{\Pi}-\I.
\end{equation}
Then, for each $j$ with $\sigma_j<1$, there exist two-dimensional
transition subspaces $\mathcal{V}_j^{\mathrm{in}}$ and
$\mathcal{V}_j^{\mathrm{out}}$ for which $U_A\mathcal{V}_j^{\mathrm{in}}=\mathcal{V}_j^{\mathrm{out}}$ and $U_A^\dagger\mathcal{V}_j^{\mathrm{out}} =\mathcal{V}_j^{\mathrm{in}}$. Thus, $W_A^{2\mathrm{ref}}$ acts invariantly on $\mathcal{V}_j^{\mathrm{in}}$ and its eigenvalues are
$e^{\pm2i\theta_j}$, where $\cos\theta_j=\sigma_j$. Explicitly,
$\mathcal{V}_j^{\mathrm{in}}=\operatorname{span}\left\{      \ket{v_j},\ket{v_j^\perp}\right\}$ and $\mathcal{V}_j^{\mathrm{out}}=\operatorname{span}\left\{       \ket{\widetilde{v}_j},\ket{\widetilde{v}_j^\perp}\right\}$,
where $\ket{v_j^\perp}=\frac{(\I-\Pi)U_A^\dagger\ket{\widetilde{v}_j}
}{\sqrt{1-\sigma_j^2}}$ and $\ket{\widetilde{v}_j^\perp}=\frac{(\I-\widetilde{\Pi})U_A\ket{v_j} }{\sqrt{1-\sigma_j^2}}$, and relative to the respective ordered bases
$\{\ket{v_j},\ket{v_j^\perp}\}$ and
$\{\ket{\widetilde{v}_j},\ket{\widetilde{v}_j^\perp}\}$, the actions of $U_A$, $U_A^\dagger$ and $W_A^{2\mathrm{ref}}$ are represented by
\begin{equation}\label{eq:singular_qubitization_matrix_reps}
    \left[U_A\right]_{
        \mathcal{V}_j^{\mathrm{in}}
        \rightarrow
        \mathcal{V}_j^{\mathrm{out}}
    }
    =
    \left[U_A^\dagger\right]_{
        \mathcal{V}_j^{\mathrm{out}}
        \rightarrow
        \mathcal{V}_j^{\mathrm{in}}
    }
    =
    \begin{pmatrix}
        \sigma_j & \sqrt{1-\sigma_j^2} \\
        \sqrt{1-\sigma_j^2} & -\sigma_j
    \end{pmatrix},
\qquad
    \left[
        W_A^{2\mathrm{ref}}
    \right]_{\mathcal{V}_j^{\mathrm{in}}}
    =
    \begin{pmatrix}
        \cos(2\theta_j) & \sin(2\theta_j) \\
        -\sin(2\theta_j) & \cos(2\theta_j)
    \end{pmatrix}.
\end{equation}

For $\sigma_j=1$, the corresponding input transition subspace reduces
to $\operatorname{span}\{\ket{v_j}\}$, on which
\begin{equation}
    W_A^{2\mathrm{ref}}\ket{v_j}
    =
    \ket{v_j}.
\end{equation}
\end{theorem}
\begin{remark}[Hermitian specialization]\label{rem:hermitian_specialization}
If $A$ is Hermitian and the projected unitary encoding satisfies
$\Pi=\widetilde{\Pi}$, the matrix identities in
\cref{eq:singular_qubitization_matrix_reps} remain valid with the singular
values $\sigma_j$ replaced by the eigenvalues $\lambda_j$. Consequently,
these identities apply to the two-reflection qubitized walk associated with
any $(\alpha,a,0)$ block encoding $U_A$ of $A$, without requiring $U_A$
itself to be Hermitian.
\end{remark}

\subsubsection{Qubitization and Chebyshev Polynomials}
\label{sec:qubitization_chebyshev_polynomials}

For the one-reflection walk operator $W_H$ associated with a $(\alpha,a,0)$ Hermitian block-encoding of a Hermitian $H$, the action on each invariant subspace $\mathcal{V}_j=\mathrm{span}\{\ket{v_j},\ket{v^{\perp}_j}\}$ is a rotation with
$\cos\theta_j=\lambda_j$, where $\ket{v_j},\ket{v^{\perp}_j}$ are as defined in \cref{thm:hermitian_qubitization}. Hence, for every positive integer $k$,
\begin{equation}
    \left[W_H^k\right]_{\mathcal{V}_j}
    =
    \begin{pmatrix}
        \cos(k\theta_j) & \sin(k\theta_j)\\
        -\sin(k\theta_j) & \cos(k\theta_j)
    \end{pmatrix}
    \nonumber\\
    =
    \begin{pmatrix}
        T_k(\cos\theta_j)
        &
        \sin\theta_j\,U_{k-1}(\cos\theta_j)
        \\
        -\sin\theta_j\,U_{k-1}(\cos\theta_j)
        &
        T_k(\cos\theta_j)
    \end{pmatrix},
    \label{eq:qubitization_chebyshev_matrix}
\end{equation}
where, $T_k$ and $U_k$ denote the Chebyshev polynomials of the first
and second kinds, respectively. Consequently,
\begin{equation}
    W_H^k\ket{v_j}
    =
    T_k(\lambda_j)\ket{v_j}
    -
    \sqrt{1-\lambda_j^2}\,
    U_{k-1}(\lambda_j)\ket{v_j^\perp}.
    \label{eq:qubitization_chebyshev_action}
\end{equation}

More generally if,
$p(x)=\sum_{k=0}^{N}a_kT_k(x)$
is a polynomial expressed in the Chebyshev basis, and $F$ is the
corresponding polynomial in the walk variable defined by $
    F(z)=\sum_{k=0}^{N}a_kz^k$, it then
follows from \cref{eq:qubitization_chebyshev_action} that
\begin{equation}
    \left[F(W_H)\right]_{\mathcal{V}_j}
    =
    \sum_{k=0}^{N}
    a_k\left[W_H^k\right]_{\mathcal{V}_j}
    =
    \begin{pmatrix}
        p(\lambda_j) & *\\
        * & *
    \end{pmatrix}.
    \label{eq:qubitized_invaraint_space_transformation}
\end{equation}
Recall that eigenvalues of $H$ are $\alpha\lambda_j$. Thus, at the operator level \cref{eq:qubitized_invaraint_space_transformation} lifts to 
\begin{equation}
    F(W_H)=    \begin{pmatrix}
        p(\frac{H}{\alpha}) & *\\
        * & *
    \end{pmatrix}.
    \label{eq:qubitized_matrix_transformation}
\end{equation}
 
For a projected unitary encoding
$(U_A,\widetilde{\Pi},\Pi)$ of a possibly non-Hermitian or rectangular matrix $A$, the resulting walk operator $W_A^{2\mathrm{ref}}$ acts on each
two-dimensional input transition subspace
$\mathcal{V}_j^{\mathrm{in}}$ as a rotation through $2\theta_j$, where $\cos\theta_j=\sigma_j$ and $\sigma_j\in[0,1)$ is a singular value of
$A/\alpha$; see \cref{thm:singular_value_qubitization}. This supports a Chebyshev construction analogous to the case above, but the doubled rotation angle and the distinct input and output transition spaces introduce additional subtleties. 

In particular,
$\left(W_A^{2\mathrm{ref}}\right)^k$ acts on
$\mathcal{V}_j^{\mathrm{in}}$ as a rotation through $2k\theta_j$. Hence, each power of the two-reflection walk operator provides a
projected unitary encoding of an even Chebyshev transform of $A$, and a polynomial $F(W_A^{2\mathrm{ref}})$ has the
corresponding linear combination of even Chebyshev transforms in its input signal block. Specifically,
\begin{equation}
\label{eq:two_reflection_even_chebyshev_transformation}
    \Pi\left(W_A^{2\mathrm{ref}}\right)^k\Pi
    =
    T_{2k}\left(\frac{A}{\alpha}\right),
    \qquad
    \Pi F\left(W_A^{2\mathrm{ref}}\right)\Pi
    =
    \sum_{k=0}^{N}a_kT_{2k}\left(\frac{A}{\alpha}\right).
\end{equation}

To obtain encodings of the corresponding odd Chebyshev transforms,
one additionally left-multiplies by $U_A$. Since $U_A$ maps the input
transition subspace $\mathcal{V}_j^{\mathrm{in}}$ to the output
transition subspace $\mathcal{V}_j^{\mathrm{out}}$, its combination
with the $2k\theta_j$ walk rotation produces the signal amplitude
$\cos\bigl((2k+1)\theta_j\bigr)=T_{2k+1}(\sigma_j)$. Consequently,
\begin{equation}
\label{eq:two_reflection_odd_chebyshev_transformation}
    \widetilde{\Pi}U_A
    \left(W_A^{2\mathrm{ref}}\right)^k\Pi
    =
    T_{2k+1}\left(\frac{A}{\alpha}\right),
    \qquad
    \widetilde{\Pi}U_AF
    \left(W_A^{2\mathrm{ref}}\right)\Pi
    =
    \sum_{k=0}^{N}
    a_kT_{2k+1}\left(\frac{A}{\alpha}\right).
\end{equation}

\begin{remark}[Singular-value interpretation]\label{rem:qubitization_chebesahv_singular_inter}
For a non-Hermitian or rectangular matrix $A$, the expressions
$T_k(A/\alpha)$ in \cref{eq:two_reflection_even_chebyshev_transformation} and \cref{eq:two_reflection_odd_chebyshev_transformation} must be understood as\textit{ singular-value transformations} rather than
ordinary matrix polynomials. The even transformations act within the
input, or right-singular-vector, space, whereas the odd
transformations map from the input space to the output, or
left-singular-vector, space. Accordingly, the projected blocks in the
two equations are input-to-input and input-to-output blocks,
respectively. These distinctions are developed formally in
\cref{sec:QSVT}.
\end{remark}

Although in
\cref{eq:qubitized_matrix_transformation,eq:two_reflection_even_chebyshev_transformation,eq:two_reflection_odd_chebyshev_transformation} the walk polynomial $F(W)$, where $W$ stands for the respective qubitized walk operator, need not itself be unitary, it can, in principle, be accessed through
an additional block or projected unitary encoding. This yields an
encoding of the corresponding polynomial transformation of
$A/\alpha$, interpreted in the singular-value sense when appropriate.
Thus, qubitization converts a Chebyshev expansion of the target
transformation into a polynomial of a unitary walk. The practical value
of this formulation depends on how efficiently the resulting walk
polynomial can be block encoded. A direct LCU construction provides
one possible route, but its efficiency may be limited by the costs of
the associated $\mathrm{PREPARE}$ and $\mathrm{SELECT}$ operations
over powers of $W$.

More direct implementations of the desired polynomial transformations
are provided by QSP~\cite{low2019hamiltonian} and
QSVT~\cite{gilyen2019quantum}, subject to the relevant boundedness and
parity conditions. The generalized QSP framework,
GQSP~\cite{motlagh2024generalized}, provides additional flexibility
for general polynomial and Laurent-polynomial transformations. We
discuss this connection in \cref{sec:gqsp}.

\subsection{Quantum Signal Processing (QSP)}\label{sec:QSP}

\begin{figure}[ht]
    \centering
    \begin{quantikz}[column sep=0.35cm]
    \lstick{$\ket{0}$}
        & \gate{R_z(\phi_d)}
        & \gate{W(x)}
        & \gate{R_z(\phi_{d-1})}
        & \cdots
        & \gate{R_z(\phi_1)}
        & \gate{W(x)}
        & \gate{R_z(\phi_0)}
        & \rstick{$P(x)\ket{0}+i\sqrt{1-x^2}Q^{\star}(x)\ket{1}$}\qw
    \end{quantikz}

     \caption{
    Degree-$d$ QSP circuit in the $W(x)$ convention. The circuit contains $d$ applications of the signal unitary $W(x)$ interleaved with $d+1$   phase rotations, and the $\ket{0}$-signal amplitude of the output is $P(x)$. }
    \label{fig:qsp}
\end{figure}
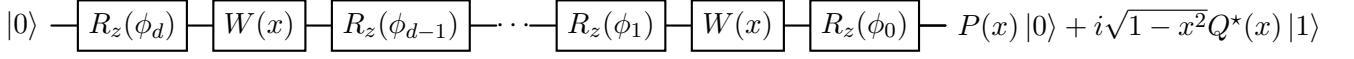

QSP provides a method for synthesizing polynomial transformations of a
scalar signal encoded in a parameter-dependent unitary in \(SU(2)\),
referred to as the \emph{signal unitary}. Several equivalent
conventions appear in the QSP literature. Here, we adopt the
\(W_x\)-convention, in which
the signal \(x\in[-1,1]\) is encoded through the \(X\)-axis rotation
\begin{equation}
\label{eq:QSP_signal_unitary}
    W(x)
    :=
    e^{i\arccos(x)X}
    =
    \begin{pmatrix}
        x & i\sqrt{1-x^2} \\
        i\sqrt{1-x^2} & x
    \end{pmatrix}.
\end{equation}
By interleaving applications of \(W(x)\) with \(Z\)-axis phase
rotations
\begin{equation}
\label{eq:QSP_phase_rotation}
    R_z(\phi)
    =
    e^{i\phi Z}
    =
    \begin{pmatrix}
        e^{i\phi} & 0 \\
        0 & e^{-i\phi}
    \end{pmatrix},
\end{equation}
one obtains a family of \(SU(2)\)-valued transformations. More
precisely, for a phase vector
$\Phi=(\phi_0,\ldots,\phi_d)\in\mathbb{R}^{d+1}$, define the
degree-$d$ QSP sequence
\begin{equation}
\label{eq:QSP_sequence}
    U_{\Phi}(x)
    =
    e^{i\phi_0Z}
    \prod_{k=1}^{d}
    \left(
        W(x)e^{i\phi_kZ}
    \right),
\end{equation}
as shown in \cref{fig:qsp}. Direct multiplication reveals that the diagonal entries of
$U_{\Phi}(x)$ are polynomials in $x$, while its off-diagonal
entries are $\sqrt{1-x^2}$ multiplied by polynomials in $x$.
Thus, the phase vector $\Phi$ determines a structured polynomial
transformation of the signal $x$.
The remarkable fact is that by choosing the phases $\phi_j$, one can implement a wide class of bounded polynomials.

\subsubsection{QSP characterization theorem}

\begin{theorem}[Ref~\cite{low2017optimal,lin2022lecture}]\label{thm:qsp_characterization} Let $p\in\R[x]$ be a real polynomial of degree $d$. Suppose that $p$ satisfies the following conditions:
    \begin{itemize}
        \item $p$ is either an even or odd polynomial.
        \item $|p(x)|\leq 1$ for all $x\in[-1,1]$.
    \end{itemize}
    Then there exist polynomials $P,Q\in\C[x]$ and angles $\Phi = (\phi_0,\dots,\phi_d)\in\R^{d+1}$ such that $p(x) = \Re(P(x))$ and
    \begin{equation}\label{eq:QSP_Equation}
        QSP_p(x) := e^{i\phi_0Z}\prod_{j=1}^d W(x)e^{i\phi_jZ} = \left[\begin{matrix}
            P(x) & iQ(x)\sqrt{1-x^2}\\
            iQ^*(x)\sqrt{1-x^2} & P^*(x)
        \end{matrix}\right].
    \end{equation}
\end{theorem}

\begin{remark}[Necessity of the constraints]
The two conditions in \cref{thm:qsp_characterization} are not only
sufficient but also necessary, although for different structural
reasons. The boundedness condition is intrinsic to encoding the
polynomial within a unitary, since every matrix entry of a unitary has
magnitude at most one. By contrast, the parity condition is an
implementation-specific artifact of the single-sequence QSP
architecture. The diagonal entries $x$ of the signal unitary $W(x)$
are odd functions of $x$, whereas its off-diagonal entries
$i\sqrt{1-x^2}$ are even. Since the phase rotations are independent of
$x$, each additional $W(x)$ layer alternates this parity structure,
forcing the output polynomial of a degree-$d$ sequence to satisfy
$p(-x)=(-1)^d p(x)$.
\end{remark}

\subsubsection{Connection with Chebyshev Polynomials and Qubitization}

Let \(H\) be Hermitian, and let \(W_H\) be the qubitized walk operator associated
with an exact \((\alpha,a,0)\) Hermitian block-encoding of \(H\).
Retain the notation
\[
    \lambda_j=\frac{\widehat{\lambda}_j}{\alpha}\in[-1,1],
\]
for the normalized eigenvalues of \(H\). On the invariant subspace
\(\mathcal{V}_j\) introduced in \cref{sec:qubitising_hermitian_block_encoding}, the qubitized walk operator acts as the rotation
\begin{equation}
    \left[W_H\right]_{\mathcal{V}_j}
    =
    \begin{pmatrix}
        \lambda_j & \sqrt{1-\lambda_j^2}\\
        -\sqrt{1-\lambda_j^2} & \lambda_j
    \end{pmatrix}.
\end{equation}
This representation is related to the QSP signal unitary
\(W(\lambda_j)\) by
\begin{equation}
    \left[W_H\right]_{\mathcal{V}_j}
    =
    e^{-i\frac{\pi}{4}Z}
    W(\lambda_j)
    e^{i\frac{\pi}{4}Z}.
\end{equation}
Consequently,
\begin{equation}
\label{eq:qsp_signal_to_qubiterate_signal}
    e^{i\widetilde{\phi}_0Z}
    \prod_{k=1}^{d}
    \left(
        \left[W_H\right]_{\mathcal{V}_j}
        e^{i\widetilde{\phi}_kZ}
    \right)
    =
    e^{i\phi_0Z}
    \prod_{k=1}^{d}
    \left(
        W(\lambda_j)e^{i\phi_kZ}
    \right),
\end{equation}
where the two phase vectors differ only at the boundary:
\begin{equation}
    \widetilde{\phi}_0=\phi_0+\frac{\pi}{4},
    \qquad
    \widetilde{\phi}_d=\phi_d-\frac{\pi}{4},
    \qquad
    \widetilde{\phi}_k=\phi_k
    \quad\text{for }0<k<d.
\end{equation}

Now given $p\in\mathbb{R}[x]$ satisfy the conditions of
\cref{thm:qsp_characterization}, let $P,Q\in\mathbb{C}[x]$ be
the corresponding polynomials, with $p=\operatorname{Re}P$.
Choosing the phase vector
$\Phi=(\phi_0,\ldots,\phi_d)$ supplied by
\cref{thm:qsp_characterization}, together with the boundary-shifted
phase vector $\widetilde{\Phi}$ defined in
\cref{eq:qsp_signal_to_qubiterate_signal}, gives
\begin{equation}
\label{eq:QSP_with_qubiterate_at_subspace_level}
    e^{i\widetilde{\phi}_0Z}
    \prod_{k=1}^{d}
    \left(
        \left[W_H\right]_{\mathcal{V}_j}
        e^{i\widetilde{\phi}_kZ}
    \right)
    =
    \begin{pmatrix}
        P(\lambda_j)
        &
        iQ(\lambda_j)\sqrt{1-\lambda_j^2}
        \\
        iQ^*(\lambda_j)\sqrt{1-\lambda_j^2}
        &
        P^*(\lambda_j)
    \end{pmatrix}.
\end{equation}

Note that, on each invariant subspace $\mathcal{V}_j$, the signal-space reflection $R$ restricts to $Z$ action. Consequently,
$e^{i\widetilde{\phi}_kR}$ restricts to the $Z$-axis phase rotation $e^{i\widetilde{\phi}_kZ}$. Replacing the scalar signal unitary $W(x)$ in the QSP sequence by $W_H$ and the associated $Z$-phase rotations by their higher-dimensional
counterparts $e^{i\widetilde{\phi}_kR}$, leads to an operator sequence that implements the action specified in
\cref{eq:QSP_with_qubiterate_at_subspace_level} on all
the invariant subspaces simultaneously. This construction is
summarized in the following theorem.

\begin{theorem}[Quantum eigenvalue transformation via qubitization \cite{low2019hamiltonian}]\label{thm:quantum_eigenvalue_transformation}
Let $U_H$ be a $(\alpha,a,0)$ block-encoding of
a Hermitian matrix $H$, and let $\Pi$, $R=2\Pi-\I$ and $W_H$ be defined as in \cref{sec:qubitising_hermitian_block_encoding}. Let $p\in\mathbb{R}[x]$ be a degree-$d$ polynomial satisfying
\begin{itemize}
    \item $p$ is either an even or an odd polynomial;
    \item $|p(x)|\leq1$ for all $x\in[-1,1]$.
\end{itemize}
Then there exists a phase vector
$\widetilde{\Phi}\in\mathbb{R}^{d+1}$ such that the corresponding
phase-modulated qubitized walk operator sequence
$U_{\widetilde{\Phi}}(W_H)$ block-encodes $P\left(\frac{H}{\alpha}\right)$. In particular,
\begin{equation}\label{eq:QSP_with_qubiterate}
    U_{\widetilde{\Phi}}(W_H):= e^{i\widetilde{\phi}_0R}
    \prod_{k=1}^{d}
    \left(
        W_H
        e^{i\widetilde{\phi}_kR}
    \right)
    =
      \begin{pmatrix}
        P\left(\frac{H}{\alpha}\right)
        &
        iQ\left(\frac{H}{\alpha}\right)
        \sqrt{\I-\left(\frac{H}{\alpha}\right)^2}
        \\[0.4em]
        iQ^*\left(\frac{H}{\alpha}\right)
        \sqrt{\I-\left(\frac{H}{\alpha}\right)^2}
        &
        P^*\left(\frac{H}{\alpha}\right)
    \end{pmatrix}.
\end{equation}
\end{theorem}

\subsubsection{Circuit Implementation for QET}
The circuit implementation of $U_{\widetilde{\Phi}}(W_H)$ requires an efficient realization of $e^{i\theta R}$ and $W_H$. Since
$W_H=RU_H$, each phase-modulated walk operator can be written in a form that allows the reflection and phase rotation to be implemented together. Relative to the decomposition
\(\operatorname{Ran}(\Pi)\oplus\operatorname{Ran}(\Pi)^\perp\) and $\widehat{\theta}=\theta+\frac{\pi}{2}$, one has

\begin{equation}
\label{eq:qubiterate_Rphi_concatenation}
e^{i\theta R}W_H
=e^{i\theta R}R U_H=\begin{pmatrix}
    e^{i\theta}\I_{\Pi} & 0 \\
    0 & -e^{-i\theta}\I_{\Pi^\perp}
\end{pmatrix}
U_H=-i
\begin{pmatrix}
    e^{i\widehat{\theta}}\I_{\Pi} & 0 \\
    0 & e^{-i\widehat{\theta}}\I_{\Pi^\perp}
\end{pmatrix}
U_H=
-i\,e^{i\widehat{\theta}R}U_H.
\end{equation}

Thus, up to a global phase, the required sequence alternates phase modulation of the signal subspace and its orthogonal complement, with applications of $U_H$. The efficiency of implementing $U_H$ depends on the chosen block-encoding construction. The required phase modulation can be implemented through a phase kickback construction using one additional ancilla qubit, as described below following \cite{gilyen2019quantum}.

Denote the additional ancilla qubit by $q$. Define the \emph{signal-controlled NOT}, denoted by $C_{\Pi}^{(q)}$, which acts on $q$ and the block-encoding register and flips $q$ precisely when the latter lies in the signal subspace.

\begin{equation}\label{eq:signal-control NOT}
C_{\Pi}^{(q)} = X_q\otimes\Pi + \I_q\otimes(\I-\Pi).
\end{equation}

Then,

\begin{equation}
\label{eq:phase_kickback_signal_reflection}
\begin{aligned}
C_{\Pi}^{(q)}
\left(e^{-i\widehat{\theta} Z_q}\otimes\I\right)
C_{\Pi}^{(q)}
\left(\ket{0}_q\otimes\ket{0^a}\right)
&=
\ket{0}_q\otimes e^{i\widehat{\theta}}\ket{0^a},
\\
C_{\Pi}^{(q)}
\left(e^{-i\widehat{\theta} Z_q}\otimes\I\right)
C_{\Pi}^{(q)}
\left(\ket{0}_q\otimes\ket{0^a}^\perp\right)
&=
\ket{0}_q\otimes e^{-i\widehat{\theta}}\ket{0^a}^\perp,
\end{aligned}
\end{equation}
and hence, applying $C_{\Pi}^{(q)}
\left(e^{-i\widehat{\theta} Z_q}\otimes\I\right)
C_{\Pi}^{(q)}$ gate combination with ancilla $q$ prepared on $\ket{0}$ results in the desired phase modulation of equation in \cref{eq:qubiterate_Rphi_concatenation}. The full circuit realization of \cref{thm:quantum_eigenvalue_transformation} is given below.

\begin{figure}[ht]
\centering
\resizebox{0.98\linewidth}{!}{%
\begin{quantikz}[row sep=0.55cm, column sep=0.20cm]
\lstick{$\ket{0}_q$}
    & \qw
    & \targ{}
    & \gate{\scriptstyle e^{-i\widehat{\phi}_0 Z_q}}
    & \targ{}
    & \qw
    & \targ{}
    & \gate{\scriptstyle e^{-i\widehat{\phi}_1 Z_q}}
    & \targ{}
    & \qw
    & \cdots
    & \targ{}
    & \gate{\scriptstyle e^{-i\widehat{\phi}_{d-1} Z_q}}
    & \targ{}
    & \qw
    & \targ{}
    & \gate{\scriptstyle e^{-i\widehat{\phi}_d Z_q}}
    & \targ{}
    & \qw
    & \rstick{$\ket{0}_q$}\qw
\\
\lstick{$\ket{0^a}$}
    & \qwbundle{a}
    & \octrl{-1}
    & \qw
    & \octrl{-1}
    & \gate[wires=2]{U_H}
    & \octrl{-1}
    & \qw
    & \octrl{-1}
    & \gate[wires=2]{U_H}
    & \cdots
    & \octrl{-1}
    & \qw
    & \octrl{-1}
    & \gate[wires=2]{U_H}
    & \octrl{-1}
    & \qw
    & \octrl{-1}
    & \meter{}
    & \rstick{$\ket{0^a}$}
\\
\lstick{$\ket{\psi}$}
    & \qwbundle{n}
    & \qw
    & \qw
    & \qw
    &
    & \qw
    & \qw
    & \qw
    &
    & \cdots
    & \qw
    & \qw
    & \qw
    &
    & \qw
    & \qw
    & \qw
    & \qw
    & \rstick{$\frac{P(H/\alpha)\ket{\psi}}{\norm{P(H/\alpha)\ket{\psi}}}$}\qw
\end{quantikz}%
}
\caption{
Ancilla-mediated phase-kickback implementation of the QET sequence
$U_{\widetilde{\Phi}}(W_H)$ from
\cref{thm:quantum_eigenvalue_transformation}~\cite{low2019hamiltonian}.
The implemented angles satisfy
$\widehat{\phi}_k=\widetilde{\phi}_k+\pi/2$ for $0\leq k<d$ and
$\widehat{\phi}_d=\widetilde{\phi}_d$. Each nonterminal phase operation
followed by $U_H$ realizes
$i\,e^{i\widetilde{\phi}_kR}W_H$. Hence, the complete circuit implements
$i^dU_{\widetilde{\Phi}}(W_H)$, differing from the desired sequence only
by the known global phase $i^d$. The open controls jointly condition on the block-encoding ancillae being in $\ket{0^a}$ and represent \emph{signal-controlled NOT} $C_{\Pi}^{(q)}$ defined in \cref{eq:signal-control NOT}, where
\(\Pi=\ket{0^a}\!\bra{0^a}\otimes I\). Each pair of $C_{\Pi}^{(q)}$, combined with \(e^{-i\phi_jZ}\), implements the
projected phase \(e^{i\phi_j(2\Pi-I)}\), while returning the
kickback qubit \(q\) coherently to \(\ket{0}\). Post-selecting the block-encoding ancillae on
$\ket{0^a}$ outputs
$\frac{P(H/\alpha)\ket{\psi}}{\norm{P(H/\alpha)\ket{\psi}}}$
with probability
$p_{\mathrm{succ}}=\norm{P(H/\alpha)\ket{\psi}}^2$ on the system register.
}
\label{fig:qet_phase_kickback}
\end{figure}
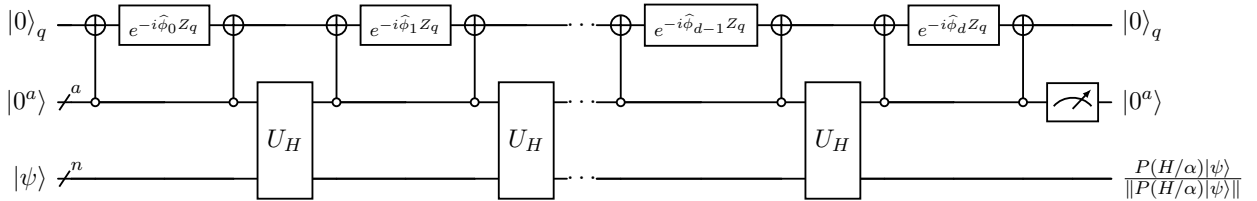
\begin{remark}[Circuit Realizations of QET]\leavevmode
\begin{enumerate}
    \item The gate-level phase-kickback circuit in
    \cref{fig:qet_phase_kickback}, which realizes the quantum
    eigenvalue transformation of
    \cref{thm:quantum_eigenvalue_transformation} for a Hermitian block
    encoding, is not presented in this form in the original
    qubitization paper. The single-ancilla construction of
    \cite[Theorem~4]{low2019hamiltonian} is instead formulated using
    controlled calls to the walk operator $W_H$ and therefore entails
    controlled access to $U_H$ when $W_H=RU_H$ is explicitly
    decomposed. The phase-kickback realization adopted here follows
    the QSVT circuit constructions of \cite{gilyen2019quantum,lin2022lecture}.

    \item A similar phase-kickback realization applies when $H$ is
    Hermitian but its block-encoding unitary $U_H$ is not. In this
    case, the sequence alternates applications of $U_H$ and
    $U_H^\dagger$ with the appropriate signal-subspace phase operators.
    This realization is outlined in the \cref{sec:QSVT}.
\end{enumerate}
\end{remark}

\subsection{Quantum Singular Value Transformation (QSVT)}

QSVT generalizes QSP from scalar signals to the singular values of a block-encoded matrix by drawing on the structural insights of both QSP and qubitization. The key step is the identification of the relevant invariant-subspace decomposition for the two-reflection qubitized walk operator
$W_A^{2\mathrm{ref}}$ associated with general projected unitary encoding $(U_A,\widetilde{\Pi},\Pi)$ of possibly non-Hermitian or rectangular $A$, as described in
\cref{sec:singular_value_qubitization}. This decomposition enables simultaneous QSP-type transformations within the corresponding two-dimensional subspaces, extending the Hermitian construction presented in the preceding section to singular-value transformations.

A QSVT of $A$ can be achieved by a sequence that alternates applications of $U_A$ and its adjoint $U_A^\dagger$ with ancilla-mediated
phase-kickbacks of the form introduced in
\cref{fig:qet_phase_kickback}, now applied with respect to the relevant input and output signal projectors of $A$. The resulting unitary has a block that is a polynomial transformation of the singular values of $A/\alpha$ where $\alpha$ is the scaling factor associated with the projected unitary encoding. As in the QSP setting, the polynomial is specified by the angles of phase rotations.

A complete derivation of QSVT is given in \cite{gilyen2019quantum}. The discussion here instead emphasizes practical understanding and structural insight. We present the main theorems underlying QSVT, highlight the subtleties introduced by non-Hermitian and rectangular matrices, provide the corresponding circuit diagrams, discuss implementation considerations, and survey representative application regimes from the literature.

\subsubsection{The Quantum Singular Value Transformation}\label{sec:QSVT}
We begin by defining what it means to have a polynomial transformation of the singular values of a matrix, especially in the context of the QSVT.
\begin{definition}[Polynomial Transformation of Singular Values]\label{def:SVT}
    Let $P$ be an odd or even polynomial and $A$ be any matrix. Let $A = W\Sigma V^\dagger$ be the SVD of $A$. By $P(\Sigma)$ we denote the diagonal matrix whose elements are $P(\Sigma)_{ii} = P(\Sigma_{ii})$. We now define the function
    \begin{equation}
        P^{(SV)}(A) :=
        \begin{cases}
             WP(\Sigma)V^\dagger, \text{ if $P$ is odd}\\
            VP(\Sigma)V^\dagger, \text{ if $P$ is even}.
        \end{cases}
    \end{equation}
\end{definition}
Given a matrix $A$ and a projected unitary encoding $(U_A,\widetilde{\Pi},\Pi)$ of $A$ with associated scaling factor $\alpha$, the goal of the QSVT is to construct a circuit which implements a projective unitary encoding of the matrix $P^{(SV)}(A/ \alpha)$. Note that from preceding definition the parity constraint is inherent in the singular-value transformation. Accordingly, when implementing QSVT, the
target polynomial $P$ must be chosen to have definite parity. 
\begin{remark}\label[remark]{remark:post_select_qsvt}
    Since the result of the QSVT is a  block-encoding of $P^{(SV)}(A/ \alpha)$, the success probability of the circuit is directly determined by the polynomial of choice, $P$. Recall (cf. \cref{thm:success_probability}) that the success probability of this block-encoding will be given by
    \begin{equation}
        P_{\text{success}} \approx \norm{P^{(SV)}(A / \alpha) \ket{\psi}}^2,
    \end{equation}
    for any state $\ket{\psi}$. Thus, if $P$ scales the singular values of $A$ down by a significant proportion, then the success probability of the circuit will suffer as a direct result. Hence, it is imperative to consider the action of $P$ on the range $[0,1]$ to ensure an optimal success probability.
\end{remark}
Next, we outline the construction of \cite{gilyen2019quantum},
which extends QSP to polynomial transformations of singular values. The central insight is to reinterpret the
qubitization--Chebyshev relations developed in
\cref{sec:qubitization_chebyshev_polynomials} within the
singular-value transformation framework of \cref{def:SVT}. As noted in
\cref{rem:qubitization_chebesahv_singular_inter}, the relations in
\cref{eq:two_reflection_even_chebyshev_transformation,eq:two_reflection_odd_chebyshev_transformation}
remain valid for a non-Hermitian or rectangular matrix $A$, provided
the resulting transformations are interpreted as the corresponding
even and odd singular-value transforms of $A$ as per \cref{def:SVT}.

To implement these transformations, one expands the two-reflection
qubitized walk operator as
\begin{equation}
    W_A^{2\mathrm{ref}}
    =
    R U_A^\dagger\widetilde{R}U_A.
\end{equation}
The resulting circuit follows the phase-kickback structure of
\cref{fig:qet_phase_kickback}, but now alternates applications of
$U_A$ and $U_A^\dagger$ with phase operators associated with the input
and output signal projectors $\Pi$ and $\widetilde{\Pi}$. As in QSP,
the phase angles are chosen to implement the target polynomial, and
the corresponding phase modulations can be combined with the
signal-space reflections already present in the walk. This yields the
standard QSVT sequence
\begin{definition}[Alternating Phase Modulation Sequence]
    Given a matrix $A$, suppose that there exists a unitary $U$ and projectors $\widetilde \Pi$ and $\Pi$ such that $A = \widetilde\Pi U \Pi$. For any degree $d\in\mathbb{Z}^+$ and any set of angles $\Phi=(\phi_1,\dots,\phi_d)\in\R^d$, we define the alternating phase modulation sequence $U_\Phi$ as follows,
     \begin{equation}
        U_\Phi := \begin{dcases}
            e^{i\phi_1(2\widetilde\Pi-\I)} U \prod_{j=1}^{\frac{d-1}{2}}\left(e^{i\phi_{2j}(2\Pi-\I)} U^\dagger e^{i\phi_{2j+1}(2\widetilde\Pi-\I)}U\right), \text{ if $d$ is odd}\\
            \prod_{j=1}^{\frac{d}{2}}\left(e^{i\phi_{2j-1}(2\Pi-\I)} U^\dagger e^{i\phi_{2j}(2\widetilde\Pi-\I)}U\right), \text{ if $d$ is even}.
        \end{dcases}
    \end{equation}
\end{definition}

The following theorem outlines all the conditions that are required for constructing the QSVT for a real-valued polynomial. We note that the conditions on the polynomial $P$ are the same as for the QSP. The added condition of the existence of projectors $\Pi$ and $\widetilde\Pi$ will boil down to the existence of a block-encoding for $A$ in practice.

\begin{theorem}[QSVT for Real Polynomials~\cite{gilyen2019quantum}]\label{thm:qsvt_real}
    Given a matrix $A$ and a degree-$d$ polynomial $P\in\mathbb{R}[x]$ which satisfies the following conditions:
    \begin{itemize}
        \item $P$ is either an even or odd polynomial.
        \item $|P(x)|\leq 1$ for all $x\in[-1,1]$.
    \end{itemize}
    Suppose that there exists a unitary $U$ and projectors $\widetilde \Pi$ and $\Pi$ such that \[A = \widetilde\Pi U \Pi.\] Then there exist angles $\Phi = (\phi_1,\dots,\phi_d)\in\R^d$ such that
    \begin{equation}
        P^{(SV)}(A) = 
        \begin{cases}
            (\bra{+}\otimes\widetilde\Pi) U^\Phi(\ket{+}\otimes\Pi), & \text{if $d$ is odd}\\
            (\bra{+}\otimes\Pi) U^\Phi(\ket{+}\otimes\Pi), & \text{if $d$ is even.}
        \end{cases}
    \end{equation}
     where
    \begin{equation}
        U^\Phi = \ketbra{0}{0}\otimes U_\Phi + \ketbra{1}{1}\otimes U_{-\Phi}.
    \end{equation}
\end{theorem}

\begin{remark}\label[remark]{remark:degree_complexity_qsvt}
    As is evident from \cref{thm:qsvt_real}, the degree of the polynomial
$P$ directly influences the circuit complexity. In particular, the
number of queries to the block-encoding unitary $U_A$ and its adjoint
scales linearly with the polynomial degree. In practice, $P$ is chosen
to approximate a target matrix function or a desired transformation
of the singular values of $A$. Although increasing the degree
generally permits a better fit of the target function, if the block encoding is approximate, its error will grow linearly with the degree (Lemma 22 of ~\cite{gilyen2019quantum}). Moreover, a higher degree may increase the numerical sensitivity
of phase synthesis. Consequently, $P$ should be
chosen to balance approximation accuracy, circuit complexity, and
numerical stability over the relevant singular-value interval.
\end{remark}

\begin{figure}[ht]
\centering

\textbf{Odd degree \(d=2\ell+1\):}\par\medskip

\resizebox{0.98\linewidth}{!}{%
\begin{quantikz}[row sep=0.30cm, column sep=0.16cm]
\lstick{$\ket{0}_q$}
  & \qw
  & \targ{}
  & \gate{e^{-i\phi_1Z}}
  & \targ{}
  & \qw
  & \targ{}
  & \gate{e^{-i\phi_2Z}}
  & \targ{}
  & \qw
  & \push{\cdots}\qw
  & \targ{}
  & \gate{e^{-i\phi_{n-1}Z}}
  & \targ{}
  & \qw
  & \targ{}
  & \gate{e^{-i\phi_nZ}}
  & \targ{}
  & \qw
  & \qw  
  & \rstick{$\ket{0}_q$}\qw
  \\
\lstick{$\ket{0}^{\otimes a}$}
  & \qwbundle{a}
  & \octrl{-1}
  & \qw
  & \octrl{-1}
  & \gate[wires=2]{U}
  & \octrl{-1}
  & \qw
  & \octrl{-1}
  & \gate[wires=2]{U^\dagger}
  & \push{\cdots}\qw
  & \octrl{-1}
  & \qw
  & \octrl{-1}
  & \gate[wires=2]{U^\dagger}
  & \octrl{-1}
  & \qw
  & \octrl{-1}
  & \gate[wires=2]{U}
  & \meter{}
  & \rstick{$\ket{0^a}$}
  \\
\lstick{$\ket{\psi}_{\mathrm{sys}}$}
  & \qwbundle{m}
  & \qw
  & \qw
  & \qw
  & \qw
  & \qw
  & \qw
  & \qw
  & \qw
  & \push{\cdots}\qw
  & \qw
  & \qw
  & \qw
  & \qw
  & \qw
  & \qw
  & \qw
  & \qw
  & \qw &\rstick{$\frac{
P^{(\mathrm{SV})}(A/\alpha)\ket{\psi}
}{
\norm{P^{(\mathrm{SV})}(A/\alpha)\ket{\psi}}
}$}\qw
\end{quantikz}
}

\bigskip

\textbf{Even degree \(d=2\ell\):}\par\medskip

\resizebox{0.98\linewidth}{!}{%
\begin{quantikz}[row sep=0.30cm, column sep=0.16cm]
\lstick{$\ket{0}_q$}
  & \qw
  & \targ{}
  & \gate{e^{-i\phi_1Z}}
  & \targ{}
  & \qw
  & \targ{}
  & \gate{e^{-i\phi_2Z}}
  & \targ{}
  & \qw
  & \push{\cdots}\qw
  & \targ{}
  & \gate{e^{-i\phi_{n-1}Z}}
  & \targ{}
  & \qw
  & \targ{}
  & \gate{e^{-i\phi_nZ}}
  & \targ{}
  & \qw
  & \qw
  & \rstick{$\ket{0}_q$}\qw
  \\
\lstick{$\ket{0}^{\otimes a}$}
  & \qwbundle{a}
  & \octrl{-1}
  & \qw
  & \octrl{-1}
  & \gate[wires=2]{U}
  & \octrl{-1}
  & \qw
  & \octrl{-1}
  & \gate[wires=2]{U^\dagger}
  & \push{\cdots}\qw
  & \octrl{-1}
  & \qw
  & \octrl{-1}
  & \gate[wires=2]{U}
  & \octrl{-1}
  & \qw
  & \octrl{-1}
  & \gate[wires=2]{U^\dagger}
  & \meter{}
  & \rstick{$\ket{0^a}$}
  \\
\lstick{$\ket{\psi}_{\mathrm{sys}}$}
  & \qwbundle{m}
  & \qw
  & \qw
  & \qw
  & \qw
  & \qw
  & \qw
  & \qw
  & \qw
  & \push{\cdots}\qw
  & \qw
  & \qw
  & \qw
  & \qw
  & \qw
  & \qw
  & \qw
  & \qw
  & \qw
  &\rstick{$\frac{P^{(\mathrm{SV})}(A/\alpha)\ket{\psi}}{\norm{P^{(\mathrm{SV})}(A/\alpha)\ket{\psi}}
}$}\qw
\end{quantikz}
}

\caption{
Explicit phase-kickback implementation of the odd- and even-degree
QSVT sequences~\cite{gilyen2019quantum}. Each pair of open-controlled NOT gates is conditioned
on the block-encoding ancillae being in \(\ket{0}^{\otimes a}\).
Together with \(e^{-i\phi_jZ}\), the pair implements the
projected phase \(e^{i\phi_j(2\Pi-I)}\), where
\(\Pi=\ket{0^a}\!\bra{0^a}\otimes I\), while returning the 
kickback qubit \(q\) coherently to \(\ket{0}\). The gates \(U\) and
\(U^\dagger\) act jointly on the block-encoding ancillae and the system
register. Post-selecting the block-encoding ancillae on
$\ket{0^a}$ outputs
$\frac{P(H/\alpha)\ket{\psi}}{\norm{P(H/\alpha)\ket{\psi}}}$
with probability
$p_{\mathrm{succ}}=\norm{P(H/\alpha)\ket{\psi}}^2$ on the system register.
}
\label{fig:qsvt-phase-kickback-parity}
\end{figure}
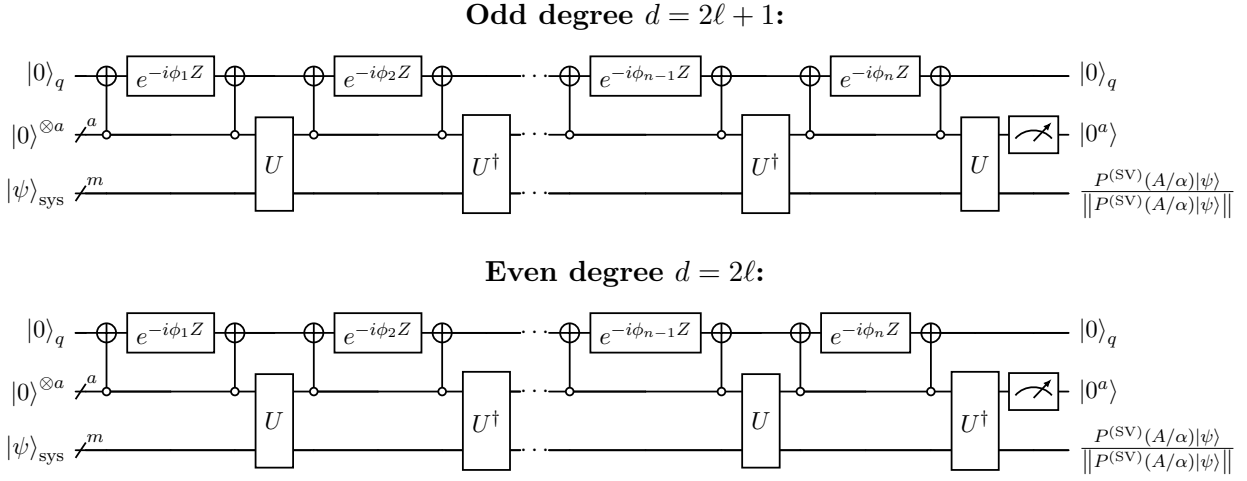
\begin{remark}[QSVT for Hermitian inputs]
Let $H$ be Hermitian and let $U_H$ be an $(\alpha,a,0)$ block encoding of
$H$, not necessarily Hermitian itself. Since the input and output signal
projectors coincide, \cref{rem:hermitian_specialization} implies that the
circuits in \cref{fig:qsvt-phase-kickback-parity} implement $P(H/\alpha)$ for any admissible polynomial $P$ of definite parity. A mixed-parity polynomial can be synthesized by coherently combining its
even and odd QSVT branches through a two-term LCU construction using one additional ancilla qubit; see \cite[Theorem~56]{gilyen2019quantum}.
\end{remark}

\subsubsection{Applying the QSVT in Practice}

Given a computational task involving a matrix function $f(A)$, one follows the following steps in order to implement the QSVT
\begin{enumerate}[label=\arabic*.]
  \item Construct an $(\alpha,a,\epsilon)$-block-encoding $U$ of $A$.
 \item Find a bounded polynomial $P$ such that $P(x)\approx g(x)$ on the relevant domain (generally this domain is $[0,1]$). 
 \item  Ensure the polynomial $P$ has definite parity and satisfies$|P(x)|\leq 1$ for all $x\in[-1,1]$.
  \item Synthesize QSP/QSVT phases for $P$.
  \item Apply the QSVT circuit using the block-encoding and its adjoint.
  \item Post-select the ancilla to be $\ket{0}$.
\end{enumerate}
Throughout the process, one should consider the post-selection probability determined by the polynomial (see \cref{remark:post_select_qsvt}) and the block-encoding $U$. It is also important to consider the degree of the polynomial, which increases the complexity of the circuit (see \cref{remark:degree_complexity_qsvt}). Furthermore, the complexity of the circuit is largely determined by the efficiency of the block-encoding, which must be carefully constructed with this in mind.

\subsubsection{Implications of the QSVT}

The QSVT framework has gained attention not just due to its ability to guide the development of new quantum algorithms but to understand and isolate exactly where quantum advantage arises in previously existing quantum algorithms. Before the introduction of the QSVT, algorithms for problems such as solving linear systems and Hamiltonian simulation all existed as independent and unrelated quantum algorithms. However, under the framework of QSVT each of these apparently unrelated algorithms can all be understood as applications of QSVT with specific choices on the block-encoding and polynomial.

For example, consider the problem of solving a system of linear equations for square $A$
\begin{equation}
    A \vec{x}=\vec{b},
\end{equation}
where we wish to prepare the state $A^{-1}\ket{\vec{b}}=\ket{\vec{x}}$, from which information about the solution $\vec{x}$ can be extracted. Note that this approach has been used to solve differential equations, computational finance and more \cite{lapworth2025precondition, wu2024finance, greenwell2026solving2dblackscholes}. The key insight into how this can be expressed as a QSVT problem is expanding $A=W\Sigma V^\dagger$ and noting that
\begin{equation}
    A^{-1}=V\Sigma^{-1}W^\dagger=V \text{diag}(1/\sigma_1,1/\sigma_2,\dots)W^\dagger.
\end{equation}
We are therefore able to choose a polynomial that approximates $P(x)=1/x$ on the domain of $\sigma_i$, and provided we have a block-encoding of $A^{\dagger}=V \Sigma W^{\dagger}$ are able to prepare the state $\ket{b}$ we will be able to generate the solution state $\ket{x}$. This is explained in more detail in \cref{example:quantum linear systems}.

Hamiltonian simulation can also be expressed as a QSVT problem. The problem of Hamiltonian simulation can be framed as beginning with an initial state $\ket{\psi}$, where we wish to simulate time evolution under a Hermitian operator $H$, that is we want to prepare the state $e^{-iHt}\ket{\psi}$. As $H$ is Hermitian its eigenvalue decomposition is $H=Q D Q^\dagger $, so $e^{-iHt}=Q e^{-iD t} Q^\dagger$. Again we see that we are able to apply QSVT to this problem as, given a block-encoding of $H$, we are able to approximate the exponential function with a polynomial in order to simulate the time evolution. Two subtleties emerge when doing this, the first being that the transformation is required on the eigenvalues rather than the singular values, which the parity condition resolves. Secondly $e^{-i t x}$ has no definite parity and so must be split into its even and odd parts. Both are treated in \cref{example:hamiltonian_simulation}. The parity constraint which necessitates the splitting and re-combining of the polynomial via LCU can be overcome by using GQSP instead of QSVT, which is discussed in the following \lcnamecref{sec:gqsp}.  

In similar ways, a wide range of quantum algorithms can be expressed in such a manner as the implementation of certain polynomial approximations to target functions of interest, with block-encodings of relevant operations \cite{martyn2021grand}. See \cref{table:qsvt_functions} for some examples.

\begin{table}[h!]
\centering
\renewcommand{\arraystretch}{1.25}
\begin{tabular}{p{0.25\linewidth}p{0.3\linewidth}p{0.35\linewidth}}
\toprule
Task & Target function & Main complexity driver \\
\midrule
Hamiltonian simulation & $e^{-i\tau x}$ & evolution time $\tau$ and precision $\epsilon$ \\
Linear systems & $1/x$ & condition number $\kappa$ \\
Spectral filtering & approximate step function & inverse spectral gap \\
Ground-state projection & low-energy filter & energy gap \\
Gibbs transformation & $e^{-\beta x}$ & inverse temperature $\beta$ \\
Amplitude amplification & polynomial in amplitude & target success probability \\
Matrix powers & $x^k$ & power $k$ and spectral domain \\
\bottomrule
\end{tabular}
\caption{Representative matrix functions implemented using QSVT.}\label{table:qsvt_functions}
\end{table}

\subsection{Generalized Quantum Signal Processing (GQSP)}\label{sec:gqsp}

Standard QSP and QSVT are extremely powerful, but their canonical formulation imposes parity restrictions on the polynomial approximation as well as requiring it to be real-valued.  In practice, these restrictions can make phase synthesis and algorithm design difficult.  GQSP aims to broaden the framework and lift these restrictions by using more general single-qubit rotations or more flexible signal-processing structures.

\subsubsection{The GQSP Theorem}\label{sec:gqsp_theorem}

In QSP, the interleaved rotations are restricted to a fixed axis, such as $Z$ rotations.  GQSP allows more general $SU(2)$ rotations of the form
\be
  R(\theta,\phi,\lambda)=
  \begin{pmatrix}
    e^{i(\phi+\lambda)}\cos(\theta) & e^{i\phi}\sin(\theta)\\
    e^{i\lambda}\sin(\theta) & -\cos(\theta)
  \end{pmatrix},
\ee
or equivalent parameterizations.  This additional freedom can remove practical restrictions on polynomial families and simplify synthesis.

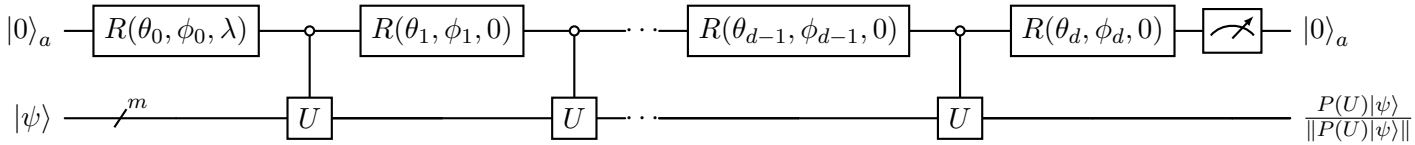
\begin{figure}[h]
\centering
\begin{quantikz}[row sep=0.55cm, column sep=0.38cm]
\lstick{$\ket{0}_a$}
  & \gate{R(\theta_0,\phi_0,\lambda)}
  & \octrl{1}
  & \gate{R(\theta_1,\phi_1,0)}
  & \octrl{1}
  & \cdots
  & \gate{R(\theta_{d-1},\phi_{d-1},0)}
  & \octrl{1}
  & \gate{R(\theta_d,\phi_d,0)}
  & \meter{}
  & \rstick{$\ket{0}_a$}
\\
\lstick{$\ket{\psi}$}
  & \qwbundle{m}
  & \gate{U}
  & \qw
  & \gate{U}
  & \cdots
  & \qw
  & \gate{U}
  & \qw
  & \qw
  & \rstick{$\frac{P(U)\ket{\psi}}{\norm{P(U)\ket{\psi}}}$}\qw
\end{quantikz}
\caption{
GQSP circuit with a zero-controlled signal unitary
$U$~\cite{motlagh2024generalized}. The upper qubit is the signal
ancilla, and each open control applies $U$ when this ancilla is in
$\ket{0}$. Post-selection on $\ket{0}_a$ outputs $\frac{P(U)\ket{\psi}}{\norm{P(U)\ket{\psi}}}$ with $p_{\mathrm{succ}}=
\norm{P(U)\ket{\psi}}^2$ on the system register.
}
\label{fig:gqsp}
\end{figure}

\begin{theorem}[Ref.~\cite{motlagh2024generalized}]\label{thm:gqsp}
    Given a unitary matrix $U$, let
    \[
        A = \ketbra{0}{0} \otimes U + \ketbra{1}{1} \otimes I.
    \]
    Let $P\in\C[x]$ be a complex polynomial of degree $d$. Suppose that 
    \[|P(e^{i\xi})|\leq 1\text{ for all }\xi\in[0,2\pi].\]
    Then there exists a polynomial $Q\in\C[x]$, a constant $\lambda\in\R$ and angles \[{\Theta =(\theta_0,\dots,\theta_d),\Phi = (\phi_0,\dots,\phi_d)\in\R^{d+1}},\] such that
    \begin{equation}
        GQSP_P(U) :=\left(\prod_{j=1}^d (R(\theta_j,\phi_j,0)\otimes I)A\right)(R(\theta_0,\phi_0,\lambda)\otimes I) = \left[\begin{matrix}
            P(U) & \cdot\\
            Q(U) & \cdot
        \end{matrix}\right],
    \end{equation}
\end{theorem}
\noindent as shown in \cref{fig:gqsp}.

As should be evident from the \lcnamecref{thm:gqsp} above, the parity restriction imposed on QSP (and thus QSVT) is lifted in the GQSP approach, thanks to the unitarity requirement on the operand $U$. Furthermore the polynomial $P$ which is applied to the matrix $U$ can have complex-valued coefficients. In comparison, QSP would require two applications of the LCU to accomplish a similar task.

\subsubsection{GQSP and Qubitization}
\label{sec:gqsp_qubitization}

QSVT leverages the mechanisms of QSP to implement polynomial
transformations of non-unitary operators supplied through block or
projected unitary encodings. GQSP, by contrast, natively implements
general complex polynomial transformations of a unitary signal.
Although this may initially appear to restrict GQSP to unitary signal
operators, block-encoding and qubitization again provides a natural bridge to extend GQSP to implement polynomial transformations of non unitary operators.

Consider first a Hermitian matrix $H$ with an $(\alpha,a,0)$ Hermitian block-encoding $U_H$, and let \(W_H\) denote the
corresponding one-reflection walk operator; see
\cref{eq:qubitization_as_rotation}. As discussed in
\cref{sec:qubitization_chebyshev_polynomials}, if
\[
    p(x)=\sum_{k=0}^{d}a_kT_k(x),
\]
is the Chebyshev expansion of a target polynomial \(p\), then setting
\(F(z)=\sum_{k=0}^{d}a_kz^k\) gives the matrix identity,
\begin{equation}
    F(W_H)
    =
    \begin{pmatrix}
        p(H/\alpha) & *\\
        * & *
    \end{pmatrix}.
\end{equation}
Consequently, applying GQSP to the signal unitary \(W_H\), with the
appropriate normalization $\gamma\geq max_|z|=1|F(z)|$, to synthesize $\frac{F(U_H)}{\gamma}$ provides a block-encoding of
\(p(H/\alpha)\).

This natural bridge between GQSP and qubitization underlies the generalized quantum eigenvalue transformation discussed in \cite{sunderhauf2023GQSVT}.

\begin{theorem}[GQET via qubitization~\cite{sunderhauf2023GQSVT}.]
\label{thm:gqet_via_qubitization}
Let \(W_H\) be the one-reflection qubitized walk operator associated with a
\((\alpha,a,0)\) Hermitian block-encoding of a Hermitian matrix \(H\).
If
\[
    p(x)=\sum_{k=0}^{d}a_kT_k(x),
    \qquad
    F(z)=\sum_{k=0}^{d}a_kz^k,
    \qquad
    \gamma\geq\max_{|z|=1}|F(z)|,
\]
then a degree-\(d\) GQSP sequence with signal unitary \(W_H\),
synthesized for \(F/\gamma\), implements an exact
\((\gamma,a+1,0)\) block-encoding of \(p(H/\alpha)\).
\end{theorem}

\subsection{Block-encoding-free GQSP for Hermitian matrices}\label{sec:gqsp_block_encoding_free}
Generalized Quantum Signal Processing (GQSP) is naturally formulated for unitary matrices, since its synthesis framework relies on sequences of single‑qubit rotations and unitary signal operators. However, in many applications the operators of interest are Hermitian but not unitary. Examples include Hamiltonians in physics, covariance matrices in statistics, and Laplacians involved in solving differential equations. In the conventional approach, a non-unitary input signal is first embedded
into a larger unitary through block encoding, after which QSP or QSVT is
applied to implement the desired polynomial transformation.
A GQSP-based alternative approach that avoids explicitly block encoding of certain Hermitian inputs is presented in \cite{mahasinghe2025hermitian}.
This approach is particularly applicable when a Hermitian matrix $H$ possesses an algebraic structure that allows it to be expressed as a real symmetric Laurent polynomial in a directly accessible unitary operator $U$. In this case an explicit block encoding of $H$ can be bypassed by using  $U$ directly as the signal operator and using GQSP to implement the corresponding Laurent-polynomial transformation of $U$ that realizes the desired polynomial transformation of $H$.

\subsubsection{The General Symmetric Laurent Access Model}
Let $U$ be an efficiently implementable primitive unitary operator acting natively on the system register. We consider a broad class of structured operators $H$ that can be modeled via a real symmetric Laurent polynomial access scheme:
\begin{equation}
H = L(U) = a_0 I + \sum_{k=1}^{t} a_k \left(U^k + (U^\dagger)^k\right), \quad a_k \in \mathbb{R}, \quad \|H\| \le 1,
\end{equation}
\noindent where $t \in \mathbb{Z}^+$ denotes the characteristic Laurent half-degree of the operator symbol. This general representation unifies several physically significant linear-algebraic primitives directly within a single framework, bypassing standard space-dilation block-encodings:

\begin{itemize}
    \item \textbf{Circulant and Structured Toeplitz Operators:} Circulant matrices showcasing strict translation invariance naturally decompose into symmetric combinations of the unitary shift operator $S$ and its adjoint $S^\dagger$. For instance, a circulant matrix with symmetric coefficients can be written as
\[
C = c_0 I + \sum_{k=1}^m c_k (S^k + S^{-k}),
\]
which is Hermitian. Since $C$ is a symmetric polynomial in $S$, polynomial transformations $P(C)$ can be
synthesized directly via GQSP applied to $S$. This covers convolution kernels, specialized structured Toeplitz operators,
and translation-invariant Hamiltonians, all without block-encoding.
    \item \textbf{Finite-Difference Laplacians:} Discretized Laplacians from PDEs provide another favorable case. For the one-dimensional heat equation
with periodic boundary conditions, the dimensionless discretized Laplacian takes the form
\[
L = S + S^\dagger - 2I,
\]
where $S$ is the periodic shift operator. Thus $L$ is a symmetric polynomial in $S$. To satisfy the contractive condition, the physical operator is rescaled as $H = L / \alpha$ where $\alpha \ge \|L\|$, and any function $f(H)$
can be implemented by synthesizing $g(x) = f(x+a_0)$ with $a_0 = \frac{-2}{\alpha}$.  This places finite-difference Laplacians squarely within the
applicability domain and enables the direct synthesis of transformations
such as diffusion propagators.
    
    \item \textbf{Szegedy-Type Quantum Walks:} The discriminant operator of a reversible Markov chain transition matrix can be expressed as $D = \frac{1}{2}(U_{\text{sz}} + U_{\text{sz}}^\dagger)$, where $U_{\text{sz}}$ is the unitary walk operator formed by a product of reflections.
\end{itemize}

\begin{remark}
The instance of the Laurent model corresponding to $t=1$,
$a_0=0$, and $a_1=1/2$, reduces the access model to the baseline identity
\begin{equation}
    H=\frac{1}{2}\left(U+U^\dagger\right),
\end{equation}
emphasized in \cite{mahasinghe2025hermitian}. Every Hermitian contraction $H$ admits such a representation with the unitary $U=H+i\sqrt{I-H^2}$. Its algorithmic applicability, depends on whether this unitary $U$ can be accessed efficiently.
\end{remark}

\subsubsection{Laurent Polynomial Composition}
Let $P(x) = \sum_{q=0}^d p_q x^q$ be a target polynomial of degree $d$ with real or complex coefficients representing the intended matrix transformation to be applied to $A$. Instead of splitting the target function into artificial symmetric branches, the algebraic structure of the access model allows us to evaluate the function as a single, unified operator-valued Laurent polynomial.

Substituting the access symbol $L(z)$ into $P(x)$ yields a compiled symmetric Laurent polynomial $C(z)$ of total half-degree $N = td$:
\begin{equation}
C(z) := P(L(z)) = b_0 + \sum_{j=1}^{N} b_j \left(z^j + z^{-j}\right) = \sum_{k=-N}^{N} b_k z^k, \quad b_k \in \mathbb{C}.
\end{equation}

\noindent where the complex Laurent coefficients $b_k$ are symmetric ($b_k = b_{-k}$) and computed strictly classically prior to circuit synthesis via the $q$-fold discrete linear convolution of the symbol's amplitudes:
\begin{equation}
b_j = b_{-j} = \sum_{q=0}^{d} p_q \left(\eta^{*q}\right)_j, \quad |j| \le td, \quad \eta_0=a_0, \; \eta_{\pm k}=a_k.
\end{equation}

\noindent Because the primitive physical operator $U$ is unitary ($U^{-1} = U^\dagger$), the operator-valued matrix function maps identically to the native Laurent form introduced by Motlagh and Wiebe \cite{motlagh2024generalized}:
\begin{equation}
P(A) = P(L(U)) = C(U) = \sum_{k=-N}^{N} b_k U^k.
\end{equation}

\noindent This classical compilation shifts the entire structural data-handling of the matrix polynomial into a single, unified coefficient string $\{b_k\}_{k=-N}^N$, matching the native operational domain of standard Laurent-GQSP without any multi-branch slicing.

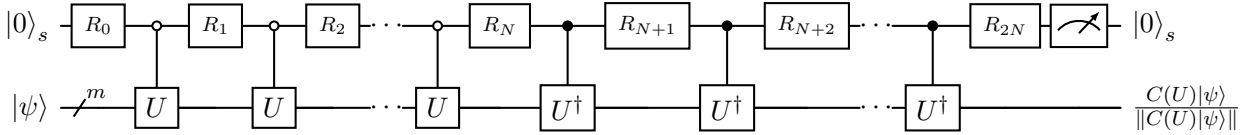
\begin{figure}[ht]
\centering
\resizebox{0.98\linewidth}{!}{%
\begin{quantikz}[row sep=0.52cm, column sep=0.18cm]
\lstick{$\ket{0}_s$}
    & \gate[
        style={
            minimum width=0.42cm,
            minimum height=0.38cm,
            inner sep=1pt
        }
      ]{\scriptstyle R_0}
    & \octrl{1}
    & \gate[
        style={
            minimum width=0.42cm,
            minimum height=0.38cm,
            inner sep=1pt
        }
      ]{\scriptstyle R_1}
    & \octrl{1}
    & \gate[
        style={
            minimum width=0.42cm,
            minimum height=0.38cm,
            inner sep=1pt
        }
      ]{\scriptstyle R_2}
    & \cdots
    & \octrl{1}
    & \gate[
        style={
            minimum width=0.44cm,
            minimum height=0.38cm,
            inner sep=1pt
        }
      ]{\scriptstyle R_N}
    & \ctrl{1}
    & \gate[
        style={
            minimum width=0.50cm,
            minimum height=0.38cm,
            inner sep=1pt
        }
      ]{\scriptstyle R_{N+1}}
    & \ctrl{1}
    & \gate[
        style={
            minimum width=0.50cm,
            minimum height=0.38cm,
            inner sep=1pt
        }
      ]{\scriptstyle R_{N+2}}
    & \cdots
    & \ctrl{1}
    & \gate[
        style={
            minimum width=0.48cm,
            minimum height=0.38cm,
            inner sep=1pt
        }
      ]{\scriptstyle R_{2N}}
    & \meter{}
    & \rstick{$\ket{0}_s$}
\\
\lstick{$\ket{\psi}$}
    & \qwbundle{m}
    & \gate{U}
    & \qw
    & \gate{U}
    & \qw
    & \cdots
    & \gate{U}
    & \qw
    & \gate{U^\dagger}
    & \qw
    & \gate{U^\dagger}
    & \qw
    & \cdots
    & \gate{U^\dagger}
    & \qw
    & \qw
    & \rstick{$\frac{C(U)\ket{\psi}}{\norm{C(U)\ket{\psi}}}$}\qw
\end{quantikz}%
}
\caption{
Implementation of $\mathcal{G}_C(U)$ following the GQSP architecture
of \cite{motlagh2024generalized}. The first $N$ signal layers apply
$U$ on the $\ket{0}_s$ branch, while the remaining $N$ layers apply
$U^\dagger$ on the $\ket{1}_s$ branch. The rotations
$\{R_j=R(\theta_j,\phi_j,\lambda_j)\}_{j=0}^{2N}$ are compiled from
the unified Laurent-coefficient sequence
$\{b_k\}_{k=-N}^{N}$. Post-selection of the signal qubit on
$\ket{0}_s$ outputs on the system regesiter with $p_{\mathrm{succ}}=\norm{C(U)\ket{\psi}}^2$.}
\label{fig:single_sequence_laurent_gqsp}
\end{figure}

To implement $C(U)$ via the Laurent-GQSP, we classically compute $2N+1$ general $SU(2)$ phase rotation angles $\{(\theta_j, \phi_j)\}_{j=0}^{2N}$ and $\lambda$. The resulting unitary is $\mathcal{G}_C(U)$:
\begin{equation}
\mathcal{G}_C(U) := \left(R(\theta_{2N}, \phi_{2N}, 0) \otimes I_s\right) A \cdots \left(R(\theta_1, \phi_1, 0) \otimes I_s\right) A \left(R(\theta_0, \phi_0, \lambda) \otimes I_s\right),
\end{equation}
\noindent where $A$ is the controlled unitary as in \cref{sec:gqsp_theorem}. The circuit diagram is given in \cref{fig:single_sequence_laurent_gqsp}. Projecting the final system state onto the ground ancilla state $\langle0|_s$ isolates the successful transformation block directly:
\begin{equation}
\left(\langle0|_s \otimes I_s\right) \mathcal{G}_C(U) \left(|0\rangle_s \otimes I_s\right) = C(U).
\end{equation}

\subsubsection{Hermitianization of non-Hermitian matrices}\label{sec:hermitianization}

In many problem cases, the matrix of interest is not Hermitian. In some cases, compared to qubitization, it may be more efficient to transform the matrix into a Hermitian analogue and then apply the block-encoding free GQSP technique. In this \lcnamecref{sec:hermitianization}, we discuss two transformations of non-Hermitian matrices into Hermitian matrices, namely via a similarity transform and Hermitian dilation.

\paragraph{Diagonal similarity transformation.}

For some non-Hermitian matrices $M$, there exists an invertible diagonal matrix $D$ such that

\begin{equation}
    H_s = D^{-1}MD,
\end{equation}
is Hermitian. Since this is a similarity transformation, $M$ and $H_s$ have the same eigenvalues, and 
\begin{equation}
    f(M) = Df(H_s)D^{-1}.
\end{equation}
For an inverse problem, this gives
\begin{equation}
    M^{-1} = DH_s^{-1}D^{-1}.
\end{equation}
The Hermitian matrix $H_s$ may therefore be processed using Hermitian-GQSP, after which the original solution is recovered using the diagonal transformations.

The main advantage of this approach is that the matrix dimension is unchanged, so no additional qubit is required for the Hermitianization itself. It also preserves the eigenvalue structure of the original matrix and therefore gives a direct relationship between $f(H_s)$ and $f(M)$. However, not every matrix is diagonally symmetrizable. The off-diagonal entries must satisfy consistency conditions, which may hold for simple one-dimensional tridiagonal discretizations but may fail for multidimensional systems with more complicated coupling structures.

A further limitation is that $D$ and $D^{-1}$ are generally non-unitary. Their implementation may require additional block-encodings, amplitude amplification or post-selection. Moreover, if the diagonal entries of $D$ vary greatly in magnitude, the condition number
\begin{equation}
    \kappa(D) = \frac{\max |d_i|}{\min |d_i|},
\end{equation}
may become large, amplifying numerical errors and reducing the success probability of the algorithm.

\paragraph{Hermitian block embedding.}

A more general method is to embed $M$ into the Hermitian matrix
\begin{equation}
    H_{\mathrm{b}}
    =
    \begin{pmatrix}
    0 & M\\
    M^\dagger & 0
    \end{pmatrix},
\end{equation}
with inverse,
\begin{equation}
H_{\mathrm{b}}^{-1}
=
\begin{pmatrix}
0 & (M^\dagger)^{-1}\\
M^{-1} & 0
\end{pmatrix}.
\end{equation}
Therefore, by preparing an input in the upper block,
\begin{equation}
\begin{pmatrix}
\ket{b}\\
0
\end{pmatrix},
\end{equation}
an odd polynomial approximation to $1/x$ applied to $H_b$ produces the desired state $M^{-1}\ket{b}$ in the lower block. The block index is represented by using one additional qubit.

The principal advantage of the block embedding is its generality. It does not require the original matrix to satisfy any symmetrizability conditions and avoids the potentially ill-conditioned diagonal matrices introduced by the similarity transform.

\addtocontents{toc}{\string\setcounter{tocdepth}{2}}

\section{Practical Guide with Example Applications}\label{sec:practical_guide}

\subsection{Comparative Scope: Constraint, Transformation and Complexity}\label{sec:comparison} 
The differences between QSVT and GQSP can be subtle and are often difficult to understand and collate. The goal of this paper is to alleviate these difficulties and illuminate the key takeaway differences between the two frameworks. To that effect, we provide three summary tables, (cf. \cref{table:comparison_polynomial_restrictions,table:comparison_circuit_complexity,table:comparison_matrix_requirements}) along with the discussion in this \lcnamecref{sec:comparison}. In addition, in the following \lcnamecrefs{sec:comparison} (\cref{sec:flowchart,example:computational_finance,example:hamiltonian_simulation,example:matrix_logarithm,example:quantum linear systems,example:spectral_filtering}) we provide a practical guide on the use of QSVT and GQSP, primarily through a flowchart (\cref{fig:flowchart}) that aims to help decide which framework to use. Along with the flowchart, we also provide several examples of practical algorithms that make use of either GQSP or QSVT, accompanied by a summary of the decisions made by following the flowchart in each example.
\begin{figure}
\begin{table}[H]
\centering
\renewcommand{\arraystretch}{1.5} 
\begin{threeparttable}
\begin{tabularx}{\textwidth}{l >{\raggedright\arraybackslash}X >{\raggedright\arraybackslash}X}
\toprule
Polynomial Restrictions &
  QSVT &
  GQSP \\ \hline
Parity &
  Polynomial must have definite parity. &
  Polynomial is not required to have definite parity. \\
Domain \& Coefficients &
  Polynomial is defined on the domain $[-1,1]$ and must have real coefficients. Polynomials with complex coefficients can be realized by splitting into real and imaginary components, and combining the result via LCU. &
  Polynomial is defined on the unit circle, and can have complex coefficients. \\
Bound &
  $|P(x)|\leq 1$ for all $x\in [-1,1]$. &
  $|P(e^{i\theta})|\leq 1$ for all $\theta\in[0,2\pi]$. \\ 
  \bottomrule
\end{tabularx}   
\end{threeparttable}

\caption{A comparison between the restrictions on the polynomial when applying QSVT and GQSP}\label{table:comparison_polynomial_restrictions}
\end{table}

\begin{table}[H]
\centering
\renewcommand{\arraystretch}{1.5} 
\begin{tabularx}{\textwidth}{l >{\raggedright\arraybackslash}X >{\raggedright\arraybackslash}X}
\toprule
Operator and Transformation &
  QSVT &
  GQSP \\ \hline
Matrix Type &
  Any matrix $A$ that can be block-encoded &
  Any unitary matrix $U$\\
Transformation &
  Polynomial transformation of the singular values of $A$ &
  Laurent polynomial transformation of $U$.  \\
Output &
  A block-encoding of $P^{(SV)}(A)$: a matrix whose singular values have been transformed by a polynomial $P$&
  The matrix polynomial $P(U)$ \\ 
  \bottomrule
\end{tabularx}
\caption{Comparison of the requirement for the operators and the mathematical transformations in QSVT and GQSP}\label{table:comparison_matrix_requirements}
\end{table}

\begin{table}[H]
\centering
\renewcommand{\arraystretch}{1.5} 
\begin{threeparttable}
\begin{tabularx}{\textwidth}{l >{\raggedright\arraybackslash}X >{\raggedright\arraybackslash}X}
\toprule
Complexity Factors &
  QSVT &
  GQSP \\ \hline
Signal queries &
   $d$ calls to $U_A$ and $U_A^\dagger$&
   $d$ calls to controlled-$U$.\\
Ancilla Requirements  &
   One ancilla qubit for QSVT, and any ancilla required for the block-encoding. If the LCU is required, an additional ancilla must be used.
   &
   One ancilla qubit. If using qubitization, then additional ancillae are required for block-encoding.\\
Additional Gates &
   Requires $d$ single-qubit $Z$-rotations and $2d$ generalized Toffoli gates acting on the QSVT ancilla.&
   Requires $d$ general $SU(2)$ .\\ 
Success Probability &
       Depends on $P$ and the singular values of $A$. Also depends on the block-encoding. See \cref{remark:post_select_qsvt}: \[
        P_{\text{success}} \approx \|{P^{(SV)}(A/\alpha) \ket{\psi}}\|^2.
    \]\vspace*{-2em}&
    Depends on $P$ and the eigenspectrum of $U$: \[
        P_{\text{success}} = \norm{P(U) \ket{\psi}}^2.
    \]\vspace*{-2.5em}
  \\
  \bottomrule
\end{tabularx}
\end{threeparttable}
\caption{A comparison between the circuit complexity, ancilla and success probability of QSVT and GQSP. Here, $P$ is a polynomial, $d$ is the degree of $P$, $U$ is a unitary and $U_A$ is a block-encoding of $A$.}\label{table:comparison_circuit_complexity}
\end{table}
\end{figure}

\subsubsection{Polynomial Conditions and Restrictions}\label{sec:comparison_poly_restrictions}
We begin the comparison by considering the conditions and restrictions required by GQSP and QSVT respectively on the polynomial functions that are to be applied (summarized in \cref{table:comparison_polynomial_restrictions}). The foremost key difference between the GQSP and QSVT is the type of polynomials which can be applied. In the case of the QSVT, a polynomial $P$ can be used if $P$ has definite parity. An important subtlety of the definite parity requirement of the QSVT is that the resulting block-encoded matrix has different singular value decompositions in the even and odd polynomial cases respectively.
\begin{remark}\label[remark]{remark:odd_even_qsvt_diff}
     If the matrix of interest, $A$, has the singular value decomposition given by $W\Sigma V^\dagger$ then an odd polynomial applied with QSVT will produce a block-encoding of the expected transformation, $WP(\Sigma)V^\dagger$. On the other hand, if $P$ is an even polynomial then the QSVT will produce a block-encoding of $VP(\Sigma)V^\dagger$, which may be an unintended consequence for certain applications.
\end{remark}

\cref{remark:odd_even_qsvt_diff} does not apply to polynomial eigenvalue
transformations in the Hermitian case when the input and output signal
spaces coincide, that is, when $\widetilde{\Pi}=\Pi$ in the projected
unitary encoding; see \cref{def:projected_unitary_encoding}. This condition
holds automatically for the standard block encoding defined in
\cref{def:block_encoding}. In this setting, the left and right singular
vectors of a Hermitian matrix can be chosen to coincide up to signs
determined by the corresponding eigenvalues. Consequently, the even and
odd singular-value transformations reduce to the corresponding polynomial
eigenvalue transformations. Thus, although a single QSVT sequence still implements only a polynomial of definite parity, a mixed-parity polynomial of the Hermitian can be synthesized by implementing its even and odd components in separate QSVT branches and combining them coherently through a two-term LCU construction using one additional ancilla qubit \cite[Theorem~56]{gilyen2019quantum}. Hence applications such as Hamiltonian simulations are unaffected by this condition. 

Since QSVT acts on singular values it requires the polynomial to have $|P(x)|\leq1$ over $x\in[-1,1]$. This ensures that the singular values of the transformed matrix at hand are not scaled beyond one. A singular value transformation with a polynomial that does not satisfy this requirement could result in a non-unitary block-encoding due to over-dilating the singular values. In the case of a polynomial $P$ with complex coefficients, one must split $P$ into two polynomials $P = \text{Re}P + i\text{Im} P$, then apply QSVT with each of the polynomials and combine the resulting block-encodings via LCU (\cref{thm:block_encoding_addition}). This adds additional cost, which must be taken into account. 

On the other hand, since GQSP acts on a unitary operator instead of singular values, admissible polynomials are not required to have a definite parity. Also since spectrum of a unitary is contained in the unit circle, a GQSP-admissible polynomial should be contractive on the unit circle. This provides greater freedom in the choice of polynomial when using GQSP. It should be noted that every polynomial which can be applied with QSVT, can also be applied with GQSP. However, the polynomial may need to be rescaled in order to fit the required bounds (within the unit circle). This rescaling will impact the success probability of the circuit, and should be taken into account when choosing between the two frameworks.

It is important to mention that the phase rotation angles determined by the polynomial at hand for QSP (and thus QSVT) were traditionally calculated by using numerical optimization methods with libraries such as pyQSP \cite{dong2021phasefactor,martyn2021grand}. This presented a large bottleneck in the QSVT workflow, for polynomials of large degree. However, the algorithm proposed by Motlagh and Wiebe in \cite{motlagh2024generalized} for GQSP provided significantly improved computational complexity, calculating the phase angles several orders of magnitude quicker. This gave GQSP a significant advantage over QSVT in practical applications where both could be applied. Recently, this phase finding algorithm was adapted to find QSP phase angles, thus bridging the gap between QSVT and GQSP in the classical computation bottleneck. The adaptation of the phase finding algorithm was done by the PennyLane team \cite{bergholm2022pennylane}, using insights from \cite{Berntson2025complementary} which describes how to transform a QSVT-ready real polynomial into a related complex polynomial.

\subsubsection{Matrix Conditions and Actions}\label{sec:matrix conditions_actions}
Our next consideration concerns the matrices to which the transformation is applied (summarized in \cref{table:comparison_matrix_requirements}). In terms of the type of matrix which can be used with QSVT and GQSP, unlike the polynomials, GQSP is more restrictive in this case. The original design of GQSP is for use with a unitary operator $U$. However, as discussed in \cref{sec:gqsp_block_encoding_free,sec:gqsp_qubitization}, it can also be extended to Hermitian matrices via qubitization or the block-encoding free access scheme described in \cite{mahasinghe2025hermitian}. On the other hand QSVT provides much more flexibility in this regard, since it can be applied to any matrix which can be block-encoded into a unitary. The difference in applicability lies in the different actions the two frameworks perform on the matrix of interest. GQSP is conceptually simpler, and acts directly on the unitary matrix $U$ as a polynomial $P$. The result, provided the circuit is successful, is a matrix which is simply $P(U)$. The QSVT is more complicated, but this difference is fundamental to how the transformation works. As the name indicates, QSVT acts on the singular values of a matrix to apply a polynomial transformation. This is termed the singular value transformation (\cref{def:SVT}). Thus, for non-Hermitian matrices, QSVT does not produce a polynomial of the matrix. The resulting matrix is a block-encoding of the singular value transformation of the matrix of interest. This fundamental distinction between the transformations implemented by the
two frameworks is central to selecting the appropriate approach in practice. It is reflected in the flowchart in \cref{fig:flowchart}, presented in \cref{sec:practical_guide}, and illustrated by the examples that follow.

\subsubsection{Circuit Complexity, Ancilla and Success Probability}
\label{sec:comparison_complexity_analysis}
The final point of comparison between GQSP and QSVT will come down to their respective circuit complexities, ancilla requirements and overall success probabilities (summarized in \cref{table:comparison_circuit_complexity}). Superficially, the circuit diagrams of QSVT and GQSP look quite similar. Both require $d$ (the degree of the chosen polynomial $P$) calls to a unitary that encodes the information required to solve the problem of interest. More specifically, QSVT will make $d$ calls to a block-encoding unitary $U_A$ and its inverse $U_A^\dagger$ (alternating between the two). On the other hand, GQSP makes $d$ calls to a controlled version of the unitary $U$. It should be noted that QSVT can be applied to unitary matrices directly, and on this factor alone it would be more efficient, requiring only $d$ calls to the unitary $U$, while GQSP requires $d$ calls to the controlled-$U$. However, these factors cannot be considered in isolation from one another. A naive comparison such as this does not take into account the polynomial restrictions, where GQSP has several advantages, as discussed in \cref{sec:comparison_poly_restrictions}. Furthermore, the circuit complexity also depends on the additional gates required for each signal processing framework. 

While in practice the circuit complexity of QSVT and GQSP is dominated by the complexity of the signal unitaries $U_A$ or $U$ respectively, there are additional rotation gates required in both circuits to realize the polynomial transformations which they perform. The QSVT circuit requires $d$ single-qubit $Z$ rotations, while the GQSP gains more flexibility in terms of the aforementioned polynomial restrictions by utilizing $d$ general $SU(2)$ rotations. In addition to the rotations, QSVT also requires $2d$ generalized Toffoli gates, which scale in size with the ancilla of the circuit. This introduces an additional scaling cost that is not present in the GQSP circuit, which must be taken into consideration when implementing the block-encoding, where there is generally a tradeoff between the ancilla cost and the gate cost. It should also be noted that the GQSP can be applied with the block-encoding free framework or with qubitization to achieve a wider variety of input matrices. These both come with additional gate cost (cf. \cref{sec:gqsp_block_encoding_free} for the former and \cref{sec:qubitization} for the latter). Qubitization also introduces the requirement of additional ancilla, which also marks a difference between QSVT and GQSP.

Both signal processing frameworks, GQSP and QSVT require an ancilla qubit to implement the phase rotations which realize the polynomial transformation. In the case of GQSP, there are no additional ancilla required, unless using qubitization to expand beyond the unitary constraints of the framework. On the other hand, since QSVT utilizes block-encodings, which by definition require ancilla qubits, there will be additional ancilla costs coming directly from the block-encoding. As noted prior, the additional ancilla will also increase the size of the generalized Toffoli gates, thus scaling the circuit complexity in addition. Therefore there is a balance to strike between the complexity of the block-encoding and the required number of ancilla. Finally, if the polynomial of interest $P$ has complex coefficients, or the matrix of interest $A$ is Hermitian and $P$ does not have definite parity, an implementation via QSVT will require the use of the LCU. Each LCU will require an additional ancilla, and will also affect the success probability of the circuit.

The success probabilities of the circuits are influenced by several factors. As noted in \cref{remark:post_select_qsvt}, the success probability of the QSVT, given a polynomial $P$, and a block-encoding $U_A$ of $A/\alpha$, acting on a state $\ket{\psi}$ is given by
\[
    P_{\text{success}} \approx \|{P^{(SV)}(A/\alpha) \ket{\psi}}\|^2.
\]
Thus the success probability is determined by the scaling factor $\alpha$, as well as the action of the polynomial on the singular values and the overlap between the state $\psi$ and the right singular vectors of $A$. We also note that the scaling factor of the polynomial $P$ will directly influence the success probability as well. Often the success probability of a circuit can be the limiting factor in implementation, especially in cases with an exponential decay in the probability. Thus it is a key factor in the consideration when choosing between GQSP, QSVT and an altogether different approach. The success probability of a GQSP circuit is conceptually simpler, since it does not depend on any scaling factor on the input operator, which is required to be unitary. In the case of the QSVT, this scaling factor on the matrix is then transformed by the polynomial $P$, which is more difficult to track. The GQSP depends only on two factors for its success probability. First, the scaling factor of $P$ which directly influences the success probability. Second, the overlap between the initial state $\ket{\psi}$ and the eigenvectors of $P(U)$. It should be noted that the block-encoding free scheme outlined in \cref{sec:gqsp_block_encoding_free} has been purported to guarantee a more stable behavior in the success probability with GQSP.

\subsection{Quantum Linear Algebra Framework and Workflow} \label{sec:flowchart}

\begin{figure}
    \center
    \includegraphics[width=1.05 \textwidth]{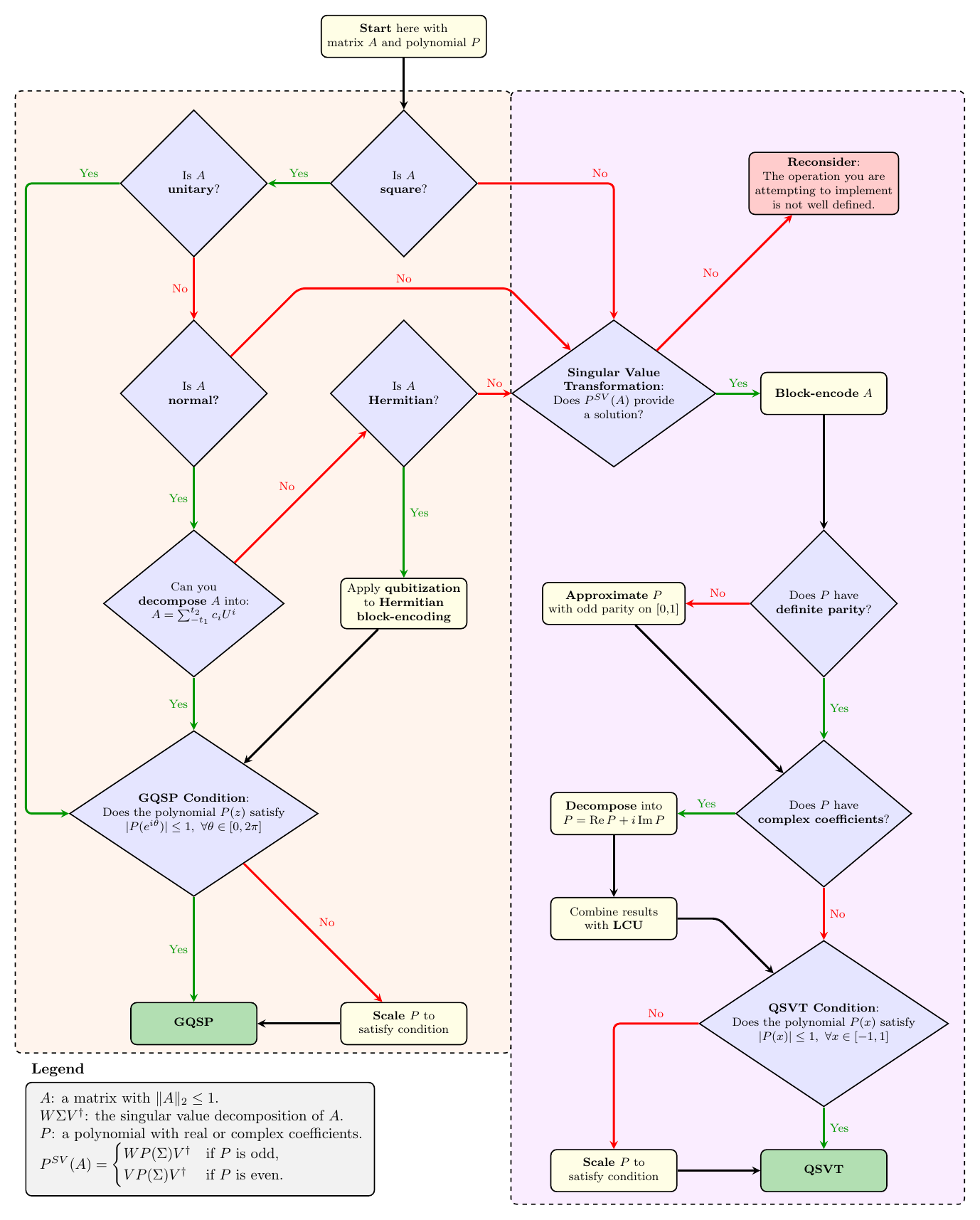}
    \caption{Implementation of matrix functions $f(A)$ using a polynomial approximation $P$.}
    \label{fig:flowchart}
\end{figure}

As the signal processing frameworks QSVT and GQSP are complex and have subtle differences in their complexity and applicability, it is easy to become lost when attempting to apply them to real world practical problems. Building upon the summary tables (\cref{table:comparison_circuit_complexity,table:comparison_matrix_requirements,table:comparison_polynomial_restrictions}) comparing QSVT and GQSP in \cref{sec:comparison}, in this \lcnamecref{sec:practical_guide} we provide a practical guide on how to best choose between the two frameworks. This guide is presented as a flowchart (\cref{fig:flowchart}) which illustrates all the questions one needs to ask when attempting to apply either QSVT or GQSP, and weighing which option is more appropriate. Starting with a matrix $A$, along with a polynomial $P$ which approximates a function $f$, to be applied to $A$, \cref{fig:flowchart} will provide a pathway for our recommended implementation strategy based on the properties of both $A$ and $P$. 
Along with the flowchart, we provide a set of example practical applications which follow the flowchart down different decision paths. The decisions made in each example on the flowchart are summarized in a blue box. In particular, examples \ref{example:spectral_filtering} and \ref{example:quantum linear systems}, on spectral filtering and quantum linear systems respectively, outline the QSVT path. Examples \ref{example:matrix_logarithm} (matrix logarithms), \ref{example:hamiltonian_simulation} (Hamiltonian simulations) and \ref{example:computational_finance} (computational finance) outline the GQSP path. Notably Example \ref{example:computational_finance} requires a reconsideration and second run through the flowchart once the matrix is Hermitianized. Meanwhile Examples \ref{example:matrix_logarithm} and \ref{example:hamiltonian_simulation} respectively outline the Laurent polynomial decomposition and qubitization pathways to apply GQSP. All of the examples are accompanied by code which can be accessed on a GitHub repository \cite{ExamplesGithub}.

\subsection{Spectral Filtering}\label{example:spectral_filtering}
A straightforward application of quantum signal processing algorithms is to produce a low-rank approximation of an image. Classically, low-rank approximations are used as a form of lossy compression. However, they can also be used to filter out noise that mostly presents as small singular values of the image. This application can be considered as a subroutine for other quantum algorithms which require filtering or thresholding of the singular values of a matrix.  

Suppose we have a greyscale image encoded as an $m$ by $n$ matrix $X$ of rank $r$, with entries $X_{ij}\in [0,1]$. Let $X = U\Sigma V^T$ be the singular value decomposition of $X$. We can obtain a rank-$k$ (with $k<r$) approximation $\widetilde X$ of $X$ by removing all but the largest $k$ singular values of $X$ from $\Sigma$. Suppose that the singular values of $X$ are $\sigma_1 \geq \sigma_2\geq\cdots\geq\sigma_r$. Without loss of generality we may assume that $\Sigma$ is the diagonal matrix $\diag(\sigma_1,\dots,\sigma_r)$.
Let $\Sigma_k$ be the diagonal matrix $\diag(\sigma_1,\dots,\sigma_k,0,\dots,0)$.
Then the low rank approximation $\widetilde X$ can be given by $\widetilde X = U\Sigma_k V^T$. 

Without prior knowledge of the singular values $\sigma_i$ it would be difficult to filter for the top $k$ values. However, we can attempt to create a low-rank approximation by filtering out all singular values below a certain threshold $\tau\in(0,1)$. Let $f$ be the thresholding function defined by
\[
    f(x) = \begin{cases}
        x,& x< -\tau \\
        0,& -\tau \leq x \leq \tau\\
        x,& x\geq \tau.
    \end{cases}
\]

\begin{figure}
    \center
\includegraphics[width=0.8 \linewidth]{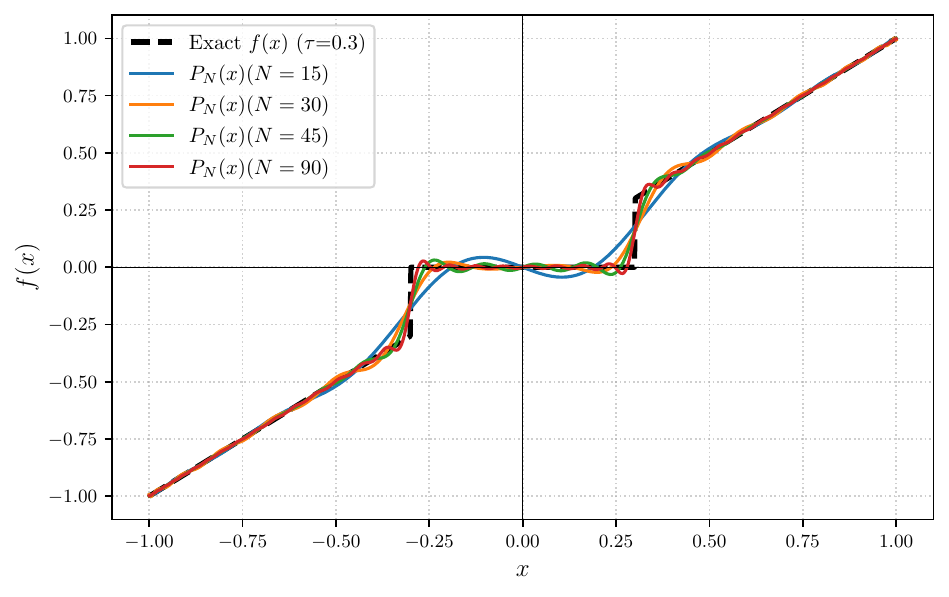}
    \caption{The approximation $p_N(x)$ of $f(x)$ for $\tau=0.3$.}
    \label{fig:step_function}
\end{figure}

Then the matrix $X_\tau = Uf(\Sigma)V^T$ will be an approximation of $X$ with lower rank. Thus our problem is to prepare $Uf(\Sigma) V^T$ using a quantum circuit, for a given $X$. Since signal processing require polynomials, we will approximate $f$ by a polynomial $p$ in the Chebyshev basis (\cref{fig:step_function}):
\begin{align}
    p_N(x) &= \frac{2}{\pi} \left[ \tau\sin(\arccos\tau) + \arccos\tau \right]T_1(x)\\
    & + \sum_{\substack{n=3\\n\text{ odd}}}^{N} \frac{2}{\pi} \left[ \frac{\sin((n-1)\arccos\tau)}{n-1} + \frac{\sin((n+1)\arccos\tau)}{n+1} \right] T_n(x),\nonumber
\end{align}
where $T_n(x)$ denotes the $n$-th Chebyshev polynomial.

We will now work through the flowchart (\cref{fig:flowchart}) to determine whether this problem is suitably posed to be solved by GQSP or QSVT.

\begin{tcolorbox}[colback=blue!5!white,colframe=blue!75!black,title=Flowchart decisions for the low-rank approximation example]
\begin{enumerate}
    \item Is the matrix square \\No, generally the matrix $A$ will not be square.
    \item Does the singular value transform $P^{SV}(A)$ provide a solution? \\Yes, we wish to apply an approximation of $f$ to the singular values of $A$.
    \item Therefore, we will need to block-encode the matrix $A$.
    \item Does $P$ have definite parity?\\ Yes, $P$ is an odd function.
    \item Does $P$ have complex coefficients?\\ No, $P$ is a real polynomial.
    \item Does $P$ satisfy the QSVT condition?\\ Yes, see \cref{fig:step_function} for an illustration of this fact.
    \begin{tcolorbox}[colback=yellow!5!white,colframe=yellow!75!black,title=QSVT condition]
    \[
        |P(x)|\leq 1,\ \forall x\in[-1,1].
    \]
    \end{tcolorbox}
    \item Therefore we can apply QSVT to obtain $UP(\Sigma)V^T$, which is a low-rank approximation of $A$.
\end{enumerate}
\end{tcolorbox}

\begin{figure}[h]
    \centering \includegraphics[width=\linewidth]{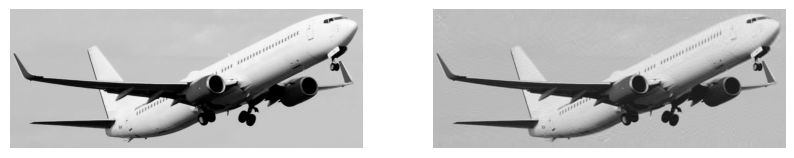}
    \caption{An example of a low-rank approximation (right) of the original image (left). Using $\tau=0.005$, the largest 67 of 281 singular values are retained.}
    \label{fig:low-rank_approx_image}
\end{figure}

\vspace{0.6cm}
It is key to notice that in this example, since the matrix $X$ is not unitary, we have to block-encode it in order to implement it within a quantum circuit. Thus the success probability of this circuit depends on the block-encoding as well as the polynomial approximation (cf. \cref{remark:post_select_qsvt}). Furthermore, the choice of block-encoding will be the main influence on the complexity of the circuit, since it will be queried $N$ times, where $N$ is the degree of the approximation $p_N$. Since $p_N$ is an approximation of the step function $f$, there is an error bound $\varepsilon \geq |f(x)-p_N(x)|$. This error is passed directly through the QSVT, to the resulting matrix. That is,
\[
    \norm{X_\tau - Up_{N}(\Sigma)V^T}\leq \varepsilon.
\]
If the block-encoding of $X$ is not exact, and has an error bound then the QSVT will also pass that error on to the resulting matrix $Up_{N}(\Sigma)V^T$ (cf. section 3.1 of \cite{gilyen2019quantum}). An example of an image processed using this low-rank approximation procedure is given in \cref{fig:low-rank_approx_image}.  

\subsection{Matrix Logarithm}\label{example:matrix_logarithm}
Given an $\epsilon_0$-accurate implementation of $U_H = e^{iH}$ for Hermitian matrix $H$ with $\lVert H\rVert<\frac{\pi}{4}$ (for example, by density matrix exponentiation as in \cref{sec:density_operators}), the matrix logarithm technique described in Corollary 71 of \cite{gilyen2019quantum} allows the implementation of a block-encoding of $H$.

First, note that $U_H$ can be transformed into a $(1, 1, \epsilon_0)$-block-encoding of
\be
\sin(H) = \frac{U_H - U_H^{\dagger}}{2i} ,
\ee
by the LCU technique, which, by the small angle approximation $\sin(z)\approx z$, can already be seen as a $(1, 1, \epsilon_0 + \lVert \sin(H) - H\rVert)$-block-encoding of $H$ itself.

If more accuracy is required, $\sin(H)$ can be further transformed into $H$ by applying a truncated Maclaurin series for $\arcsin$:
\be
P(x) = \sum_{k=0}^{d}
\frac{(2k)!}{4^k (k!)^2 (2k+1)}\,x^{2k+1} \approx \arcsin(x).
\ee
\begin{figure}[h]
    \centering
    \includegraphics[width=0.8 \linewidth]{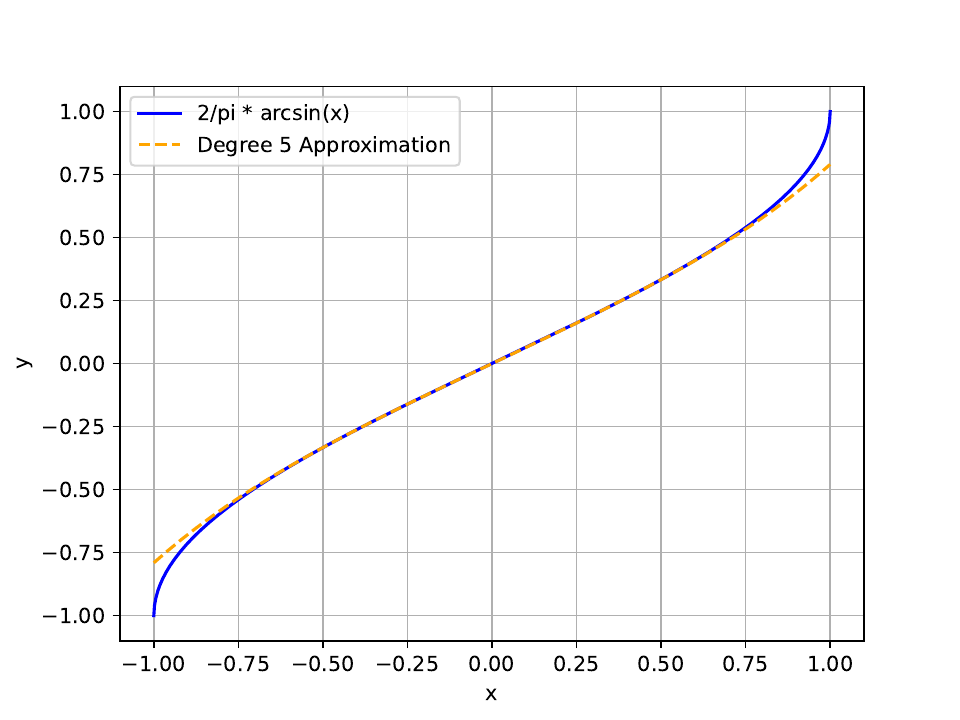}
    \caption{Comparison of $\frac{2}{\pi}\arcsin$ to its Maclaurin series truncated at degree 5.}
    \label{fig:arcsin_approximation}
\end{figure}
Scaling to $\frac{2}{\pi} P(x)$ ensures the resulting polynomial satisfies the QSVT condition of being bounded by 1 on the domain $[-1, 1]$ (see \cref{fig:arcsin_approximation}). The authors of \cite{gilyen2019quantum} then apply QSVT to the block-encoded $\sin(H)$, resulting in a $(1, 2, \epsilon_1 + \epsilon_2)$-block-encoding of
\be
\frac{2}{\pi} \arcsin(\sin(H)) = \frac{2}{\pi} H ,
\ee where $\epsilon_1$ is the error resulting from propagating the error $\epsilon_0$ through the QSVT circuit, and $\epsilon_2$ is the error resulting from the truncation of the $\arcsin$ series. For any fixed $\lVert H\rVert < \arcsin(1) = \frac{\pi}{4}$, the degree $d$ required to obtain an error lower than $\epsilon_2$ for the $\arcsin$ approximation on the spectrum of $\sin(H)$, is $O(\log(1/\epsilon_2))$.  By the robustness of the singular value transformation
\cite[Lemma~22]{gilyen2019quantum}, $\epsilon_1 \leq 4d\sqrt{\epsilon_0}$, which increases linearly with the degree, while $\epsilon_2$ decreases exponentially. In summary, ensuring for example $\lVert H\rVert=\frac{1}{2}$ by scaling, a $(\pi, 2, \epsilon)$-block-encoding of $H$ is obtained, with $\epsilon =\pi(\epsilon_1 + \epsilon_2)$. 

However, following the flowchart in \cref{fig:flowchart}, a more efficient route is indicated. The same transformation can be applied directly on $U_{H}$, by composing the polynomial $\frac{2}{\pi} P(x)$ with the Laurent polynomial $L(z) = \frac{z - z^{-1}}{2i}$ resulting in Laurent polynomial $P' = \frac{2}{\pi}P(\frac{z - z^{-1}}{2i})$, which can be applied to $U_{\rho t}$ using GQSP. This composition avoids the LCU block-encoding, saving one ancilla and providing a $(\pi, 1, \epsilon)$-block-encoding of $H$.

\begin{tcolorbox}[colback=blue!5!white,colframe=blue!75!black,title=Flowchart decisions for the matrix logarithm example]
\begin{enumerate}
    \item Is $A = \sin(H)$ square? \\ Yes.
    \item Is $A = \sin(H)$ unitary?\\ No, generally $A$ will not be a unitary matrix.
    \item Is $A = \sin(H)$ normal? \\Yes, it is even Hermitian.
    \item Can you
decompose $A = \sin(H)$ into:
$A = \sum^{t_2}_{-t_1} c_i U^i$  \\ Yes, $A$ is a Laurent polynomial of $U_H$: $A = L(U_H)= \frac{1}{2i} U_H + \frac{1}{2i} U_H^{-1}$. 
    \item Does $P' = \frac{2}{\pi} P \circ L$ satisfy the GQSVT condition?\\ Yes, $L$ maps the unit circle to the interval $[-1, 1]$ (as it acts as the $\sin$ function on the phase of the input) and $P$ is bounded by $\frac{\pi}{2}$ on the interval $[-1, 1]$.
    \begin{tcolorbox}[colback=yellow!5!white,colframe=yellow!75!black,title=GQSP Condition]
    \[
        |P'(e^{i \theta})|\leq 1,\ \forall \theta \in[0,2\pi].
    \]
    \end{tcolorbox}
    \item Therefore we can apply $\frac{2}{\pi}P$ to $A$ by applying $P'$ to $U_H$ using GQSP.
\end{enumerate}
\end{tcolorbox}
 
\subsection{Quantum Linear Systems}
\label{example:quantum linear systems}
The quantum linear systems problem asks to prepare a state proportional to $\ket{x}\in\mathbb{C}^n$ such that
\be
 A\ket{x}=\ket{b},
\ee
for known matrix $A\in\mathbb{C}^{m\times n}, ||A||_2\le 1$ and $\ket{b}\in \mathbb{C}^m$. From the singular value decomposition of $A=U\Sigma V^\dagger$, then its adjoint is $A^\dagger=V \Sigma U^\dagger$ and Moore-Penrose pseudoinverse is $A^+=V\Sigma^+U^\dagger$. This suggests for the odd function $f$ such that,
\be
 f(x)= \begin{cases}
     \frac{1}{x}, & x\ne 0 \\
     0, & x=0,
 \end{cases}
\ee
then we have,
\be
f^{SV}(A^\dagger)=A^+.
\ee
When applying this in practice we only approximate $\frac{1}{x}$ within the range of the singular values of $A$,
\be
 P(x)\approx \begin{cases}
     \frac{1}{x}, & x\in[-1,-1/\kappa]\cup[1/\kappa,1] \\
     0, & x=0,
 \end{cases}
\ee
where $\kappa=\frac{\sigma_{max}}{\sigma_{min}}$ is the condition number of $A$. The polynomial degree scales with $\kappa$ and the desired precision.  QSVT-based linear-system algorithms provide a clean and often optimal formulation of this task \citep{harrow2009quantum,childs2017quantum,gilyen2019quantum}.

\begin{tcolorbox}[colback=blue!5!white,colframe=blue!75!black,title=Flowchart decisions for the quantum linear systems example]
\begin{enumerate}
    \item Is the matrix square? \\No, generally the matrix $A$ will not be square.
    \item Does the function need to be applied to the singular values of $A$? \\Yes, we wish to apply the polynomial $P$ to the singular values of $A$.
    \item We find a block-encoding of $A$
    \item Does $P$ have definite parity?\\ Yes, $P$ is an odd polynomial.
    \item Does the $P$ have complex coefficients?\\ No, $P$ is a real polynomial.
    \item Does $P$ satisfy the QSVT condition?\\
    No, $P$ needs to be rescaled depending on $\kappa$ and the approximation used.
    \item Therefore we can apply the QSVT to get $A^+$.
    \begin{tcolorbox}[colback=yellow!5!white,colframe=yellow!75!black,title=QSVT condition]
    \[
        |P(x)|\leq 1,\ \forall x\in[-1,1].
    \]
    \end{tcolorbox} 
\end{enumerate}
\end{tcolorbox}

\subsection{Hamiltonian Simulation}
\label{example:hamiltonian_simulation}
The goal of Hamiltonian simulation is to efficiently implement the unitary operator $e^{-itH}$ for a given Hamiltonian $H$ and time $t$, up to an accuracy of $\epsilon$. As $H$ is Hermitian it will have a complete set of eigenvectors $\ket{\psi_j}$ with corresponding eigenvalues $\hat{\lambda}_j$.

To do this we assume we have an efficient way of implementing a block-encoding of the Hamiltonian $H$, a requirement for both QSVT and GQSP. In general we may need to rescale our Hermitian matrix $H$ by some $\alpha > ||H||_2$
\begin{equation}
    H \to H/\alpha \quad \lambda_j = \hat{\lambda}_j/\alpha \quad \tau = \alpha t \quad  e^{-i t \hat{\lambda}_j} = e^{-i \tau{\lambda}_j}.
\end{equation}
The GQSP protocol will implement polynomials of a unitary matrix, which for our block-encoded $U$ may not necessarily carry physical meaning. This leads us to introduce a qubitization walk operator as explained in \cref{sec:qubitization}, so that repeated action forms meaningful polynomials of the eigenvalues of $H$. Qubitizing as per \cref{eq:qubitization_walk_operator} will give the operator $W=RU$ such that for $\ket{v_j}=\ket{0^a}\ket{\psi_j}$
\begin{equation}
    W^k \ket{v_j} = \cos(k\theta_j)\ket{v_j} - \sin(k\theta_j)\ket{v_j^{\perp}},
    \qquad \cos\theta_j = \lambda_j,
\end{equation}
so that the component retained on post-selection is
$\cos(k\theta_j) = T_{|k|}(\lambda_j)$, where we have set $\cos{\theta_j}=\lambda_j$, and $\ket{v_j^\perp}$ as the orthogonal component to $\ket{v_j}$. Upon post selection of the ancilla on $\ket{0^a}$, we will recover the transformed state $T_{|k|}(\lambda_j)\ket{\psi_j}$. The goal is now to find coefficients $a_k$ such that
\begin{equation}
    \sum_{k=0}^da_kT_k(x)\approx e^{-i\tau x},
    \label{eq:exponential expansion}
\end{equation}
and then to realize them as a polynomial that is bounded on the unit circle. This can be thought of as expressing $e^{-i \tau x}$ in the Chebyshev basis, and can be found through the Jacobi-Anger expansion, as
\begin{equation}
    c_0=J_0(\tau), \qquad a_k=2(-i)^kJ_k(\tau),
\end{equation}
where $J_k$ denotes the Bessel Functions of the first kind. Simply choosing the polynomial $\sum_{k=0}^d a_kz^k$ would correctly result in the required action on the eigenstates $\ket{v_j}$ of $W$, but this polynomial will not automatically satisfy the GQSP unitary constraint. However, as both $z^k$ and $z^{-k}$ map to $T_k(x)$ in terms of the induced action on $\lambda_j$ we are able to use this freedom to choose a symmetric Laurent series $\tilde{P}(z)=\sum_{k=-d}^dp_kz^k$ to satisfy the GQSP condition. By taking 
\[a_0=c_0 \text{ and } \,p_k=p_{-k}=\frac{1}{2}a_k\, \text{ for } k \geq 1,\]
we  will satisfy the GQSP condition as follows, 
\begin{equation}
    \tilde P(e^{i\theta}) = \sum_{k=-d}^{d}p_ke^{ik\theta}
= a_0 + \sum_{k=1}^{d}p_k\big(e^{ik\theta}+e^{-ik\theta}\big)
= \sum_{k=0}^{d}a_kT_k(\cos\theta) 
\approx e^{-i \tau \cos{\theta}}.
\end{equation}
where the last step comes from our choice of $a_k$ to match the $e^{-i\tau x}$ function up to some error $\epsilon$. For all $\theta\in [0,2\pi]$, we see $|\tilde{P}(e^{i\theta})| \leq 1+\epsilon$, which can be rescaled by $1+\epsilon$ to fit the GQSP condition of being bounded by 1. Noting now we can form the polynomial, (rather than the Laurent polynomial), into
\begin{equation}
    \tilde{P}(z)=z^{-d}\sum_{k=0}^{2d}a_{k-d}z^k=z^{-d}P(z),
\end{equation}
we can implement this transformation with $d$ applications of $W^{\dagger}$ followed by the GQSP protocol with the polynomial $P(z)$, which will also satisfy the GQSP conditions as for $z \in S_1$, $|P(z)|=|z^d\tilde{P}(z)|\leq1$.
Post-selecting both the GQSP ancilla and the block-encoding register on
$\ket{0}$ and $\ket{0^a}$ respectively, the surviving amplitude on each
$\ket{v_j}$ is $\tilde{P}(e^{i\theta_j})\approx e^{-i\tau\lambda_j}$, so that
for an arbitrary input $\ket{\psi}=\sum_j\gamma_j\ket{\psi_j}$ the circuit
prepares
\begin{equation}
    \sum_j \gamma_j e^{-i\tau\lambda_j}\ket{\psi_j}
    = e^{-iHt}\ket{\psi},
\end{equation}
to accuracy $\epsilon$. The fast decay of the coefficients generated by the Bessel functions results in the GQSP-based Hamiltonian simulation achieving query complexity essentially
\begin{equation}
  O\left(\alpha t + \frac{\log(1/\epsilon)}{\log\log(1/\epsilon)}\right),
\end{equation}
matching lower bounds in important oracle models \citep{low2017optimal,low2019hamiltonian}. The success probability of post-selection on the GQSP ancilla state $\ket{0}$ and the block-encoding ancilla state $\ket{0^a}$ is close to 1, since the resulting block-encoding approximates the unitary evolution operator $e^{-iHt}$ up to error $\epsilon$. As the target evolution is unitary, deviations from the desired block correspond only to a small leakage into the orthogonal ancilla subspace.

\begin{tcolorbox}[colback=blue!5!white,colframe=blue!75!black,title=Flowchart decisions for the Hamiltonian simulation example]
\begin{enumerate}
    \item Is $H$ square? \\ Yes, as it is Hermitian.
    \item Is $H$ unitary? \\
    No, $H$ is not a unitary matrix.
    \item Is $H$ normal? \\Yes, as it is Hermitian.
    \item Can you
decompose H into:
$H = \sum^{t_2}_{-t_1} c_i U^i$  \\The problem Hamiltonian is not inherently in this form. 
    \item Is $H$ Hermitian? \\ Yes.
    \item Qubitize the block-encoding of $H$.
    \item Does $P$ satisfy the GQSP condition?\\ No, the naive polynomial does not.
    \item Scale $P$ to satisfy condition \\ Done through Laurent series expansion.
    \begin{tcolorbox}[colback=yellow!5!white,colframe=yellow!75!black,title=GQSP Condition]
    \[
        |P(e^{i \theta})|\leq 1,\ \forall \theta \in[0,2\pi].
    \]
    \end{tcolorbox}
    \item Therefore we can apply the GQSP.
\end{enumerate}
\end{tcolorbox}

\begin{figure}[h]
    \centering
    \includegraphics[width=1\linewidth]{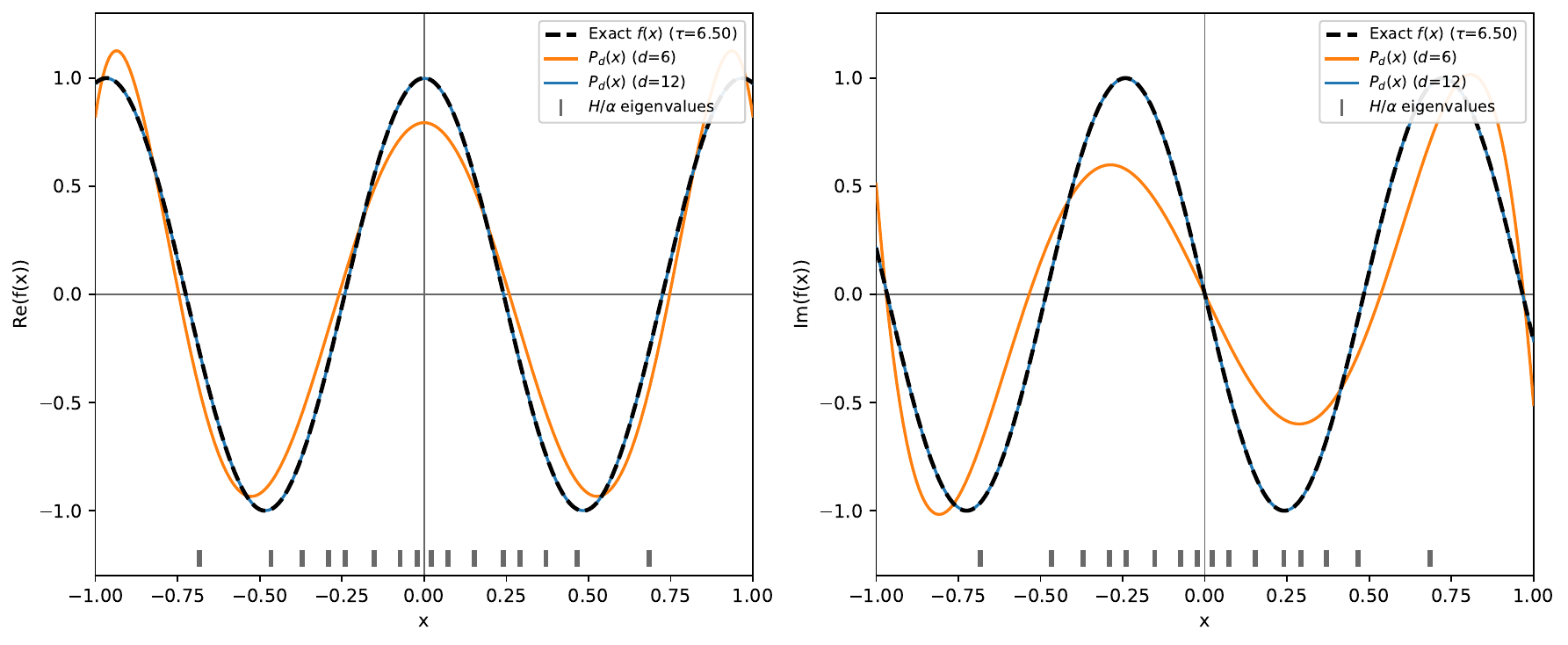}
    \caption{Real and imaginary parts of the truncated Chebyshev/Jacobi--Anger polynomial $P_d(x)$, compared against the exact target $e^{-i\tau x}$. The eigenvalues of $H/\alpha$ are also shown, as these are the only points at which the approximation quality is relevant to the simulation.}
\label{fig:GQSPcoefficientmatching}
\end{figure}

 We now show the implementation of this for a specific Hamiltonian, chosen to be the open-chain transverse-field Ising Model:
 \begin{equation}
 H = -J\sum_{i=0}^{3} Z_iZ_{i+1} \;-\; h\sum_{i=0}^{4} X_i, \qquad J=1.0,\ h=0.5,\ t=1.0,\ \epsilon=10^{-3}.
 \label{eq:TFIM}
 \end{equation}
 The $ZZ$ and $X$ terms correspond to Pauli operations on the system. As we are only considering 5 qubits the system is small enough to verify by simulation by explicitly computing $e^{-iHt}$ and acting on the initial state which we take to be $\ket{00000}$. With the parameters of $H$ given above we choose $\alpha=6.5$ to be rescaled for the qubitiztaiton and block encoding. To get a Chebyshev approximation to $e^{-i \tau x}$ as per \cref{eq:exponential expansion} within our error $\epsilon=10^{-3}$. As shown in \cref{fig:GQSPcoefficientmatching} a $d=12$ degree polynomial to satisfy this, where we have also plotted the eigenvalues of $H/\alpha$ as it is at these points which we wish our approximation to be valid at. The phase factors corresponding to this polynomial are found using standard GQSP angle-finding via Fejér–Riesz completion.

\begin{figure}[h]
     \centering
     \includegraphics[width=0.6\linewidth]{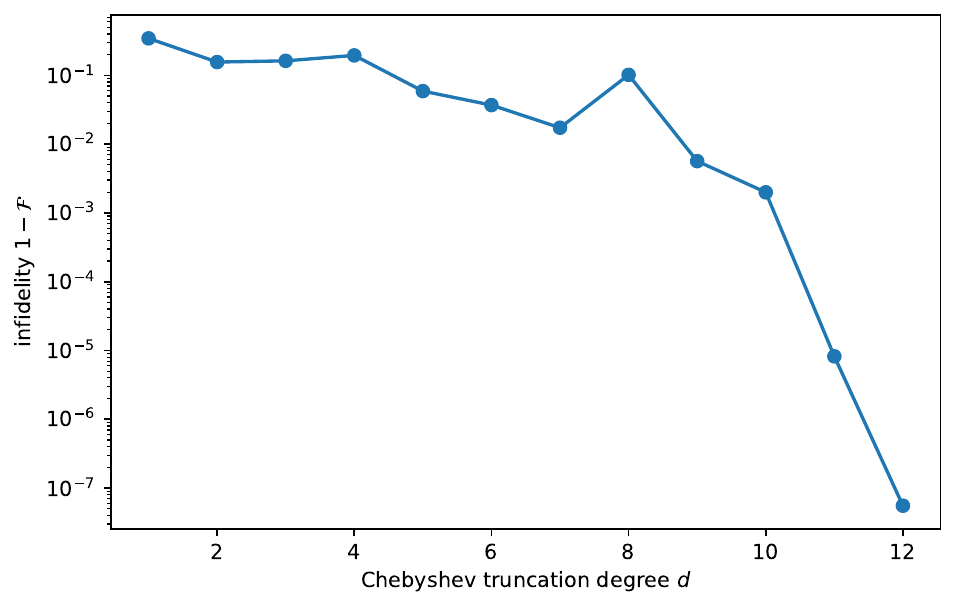}
     \caption{The infidelity of the final state produced by the GQSP Hamiltonian simulation for 5 qubits as specified in Eq. \ref{eq:TFIM} compared to the true final state.}
     \label{fig:GQSPfidelitytotargetstate}
 \end{figure}

 Implementing this into the GQSP circuit construction gives the following results for infidelity to the true target state, as shown in \cref{fig:GQSPfidelitytotargetstate}. Note that the final infidelity is much lower than our initial error $\epsilon=10^{-3}$ as this is only the polynomial approximation error and the actual evolution of the eigenvalues may be much closer to the desired behaviors than this lower bound.

\subsection{Differential Equations}\label{example:computational_finance}

Discretized partial differential equations, such as the Black--Scholes equation, lead to linear systems and matrix-function problems.  Block-encoding and QSVT/GQSP can be used to implement time-step inverses, resolvents, or polynomial approximations to finite-difference operators.

In the one-dimensional case, the Black--Scholes equation for a European option is 
\begin{equation}
    \frac{\partial V}{\partial t} + \frac{1}{2}\sigma(S,t)^2 S^2 \frac{\partial^2 V}{\partial S^2} +rS \frac{\partial V}{\partial S} -rV = 0.
\end{equation}
This can be discretized, resulting in a linear systems problem of the form
\begin{equation}
    V^{t-\Delta \tau} = \tilde{M}^{-1}V^t.
\end{equation}
With the known initial condition $V^t$, $\tilde{M} := \mathbb{I} + M$, and
\begin{equation}
M
=
\Delta\tau
\begin{bmatrix}
r & -\epsilon & 0 & 0 & \cdots & 0\\
-\gamma_1 & \gamma_1+\beta_1+r & -\beta_1 & 0 & \cdots & 0\\
0 & -\gamma_2 & \gamma_2+\beta_2+r & -\beta_2 & \cdots & 0\\
0 & 0 & -\gamma_3 & \ddots & \ddots & \vdots\\
\vdots & \vdots & \vdots & \ddots & \ddots & -\beta_{2^n-2}\\
0 & 0 & 0 & \cdots & -\epsilon & 0
\end{bmatrix},
\qquad 0 <\epsilon <<1,
\end{equation}
We adopt the time-independent local-volatility model
\begin{equation}
\sigma(S,t)=\frac{\nu}{\sqrt{S}}, \qquad S>0, \qquad \nu > 0.
\end{equation}
Thus, at each interior grid point \(S_j>0\),
\(\sigma_j:=\sigma(S_j,t)=\nu/\sqrt{S_j}\).
The finite-difference coefficients \(\gamma_j\) and \(\beta_j\) are
defined as follows:
\begin{equation}
\gamma_j
=
\frac{\sigma_j^2 S_j^2}{(S_j-S_{j-1})(S_{j+1}-S_{j-1})}
-\frac{rS_j}{S_j-S_{j-1}}
=
\frac{\nu^2 S_j}{(S_j-S_{j-1})(S_{j+1}-S_{j-1})}
-\frac{rS_j}{S_j-S_{j-1}},
\end{equation}
and
\begin{equation}
\beta_j
=
\frac{\sigma_j^2 S_j^2}{(S_{j+1}-S_j)(S_{j+1}-S_{j-1})}
=
\frac{\nu^2 S_j}{(S_{j+1}-S_j)(S_{j+1}-S_{j-1})}.
\end{equation}

Thus, the problem reduces to finding the inverse of $\tilde{M}$. Following the diagram of \cref{fig:flowchart}, the singular value transformation is suggested, because $\tilde{M}$ is not Hermitian. Indeed, the inverse $\tilde{M}$ can be implemented by approximately applying the function $\frac{1}{x}$ to the singular values of $\tilde{M}$, but for multiple (say $k$) steps at once, we need to implement $\tilde{M}^{-k}$, which is not the same as a singular value transformation by $\frac{1}{x^k}$. To avoid having to repeat a costly QSVT circuit $k$ times, and avoid the corresponding diminishing success probability, we answer ``No" to the question ``Does the singular value transformation provide a solution", and we reconsider this problem.  

\subsubsection{Hermitianization via similarity transformation}

\begin{figure}
    \centering

    \begin{subfigure}[h]{0.70\textwidth}
        \centering
        \includegraphics[width=\linewidth]
            {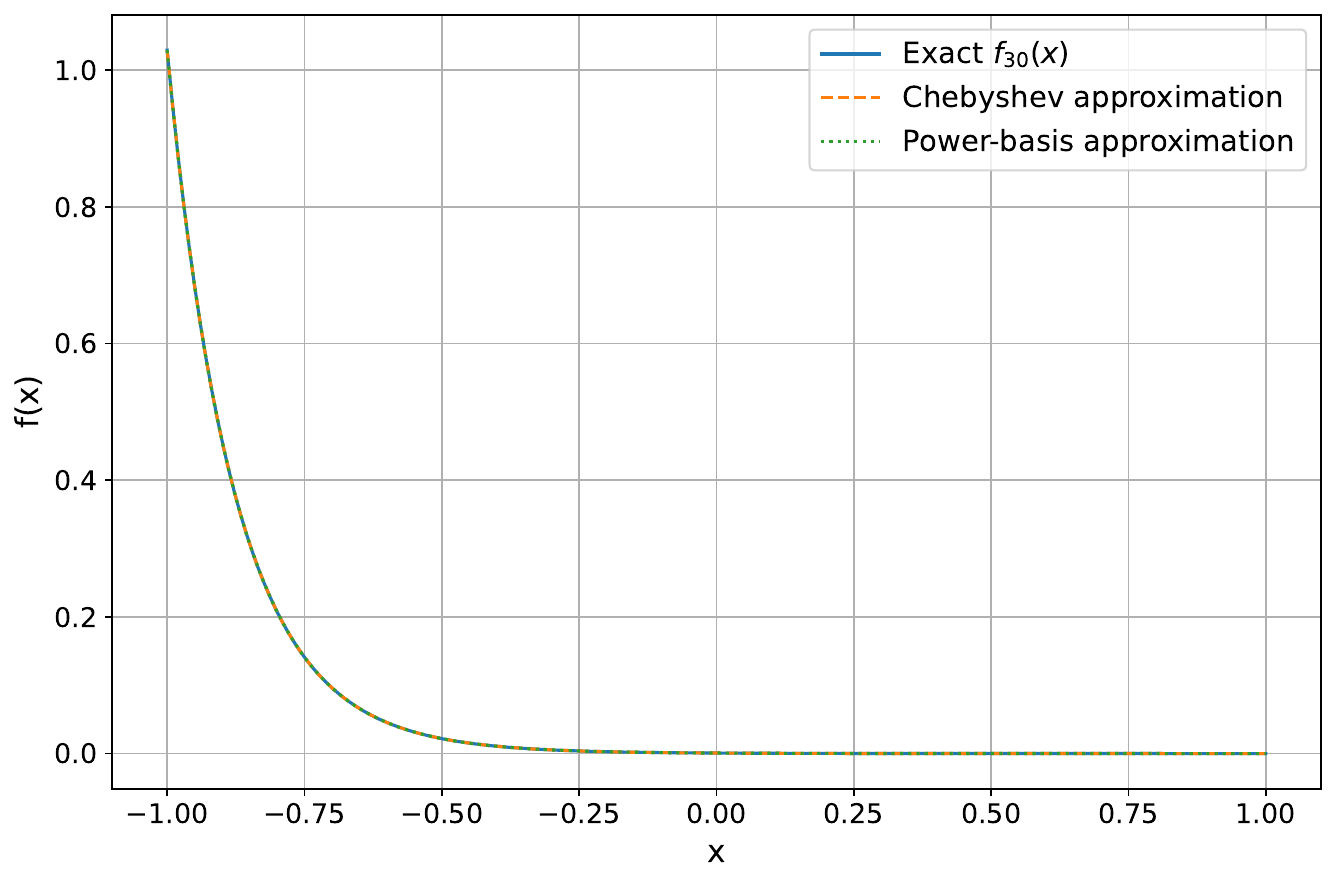}
        \caption{
            Comparison of the exact function \(f_{30}(x)\) with its degree-20
            Chebyshev and power-basis polynomial approximations over
            \(x \in [-1,1]\). Both polynomial representations closely track the
            exact function across the entire interval.
        }
        \label{fig:finance_poly_approx}
    \end{subfigure}

    \vspace{0.8cm}

    \begin{subfigure}[h]{0.70\textwidth}
        \centering
        \includegraphics[width=\linewidth]
            {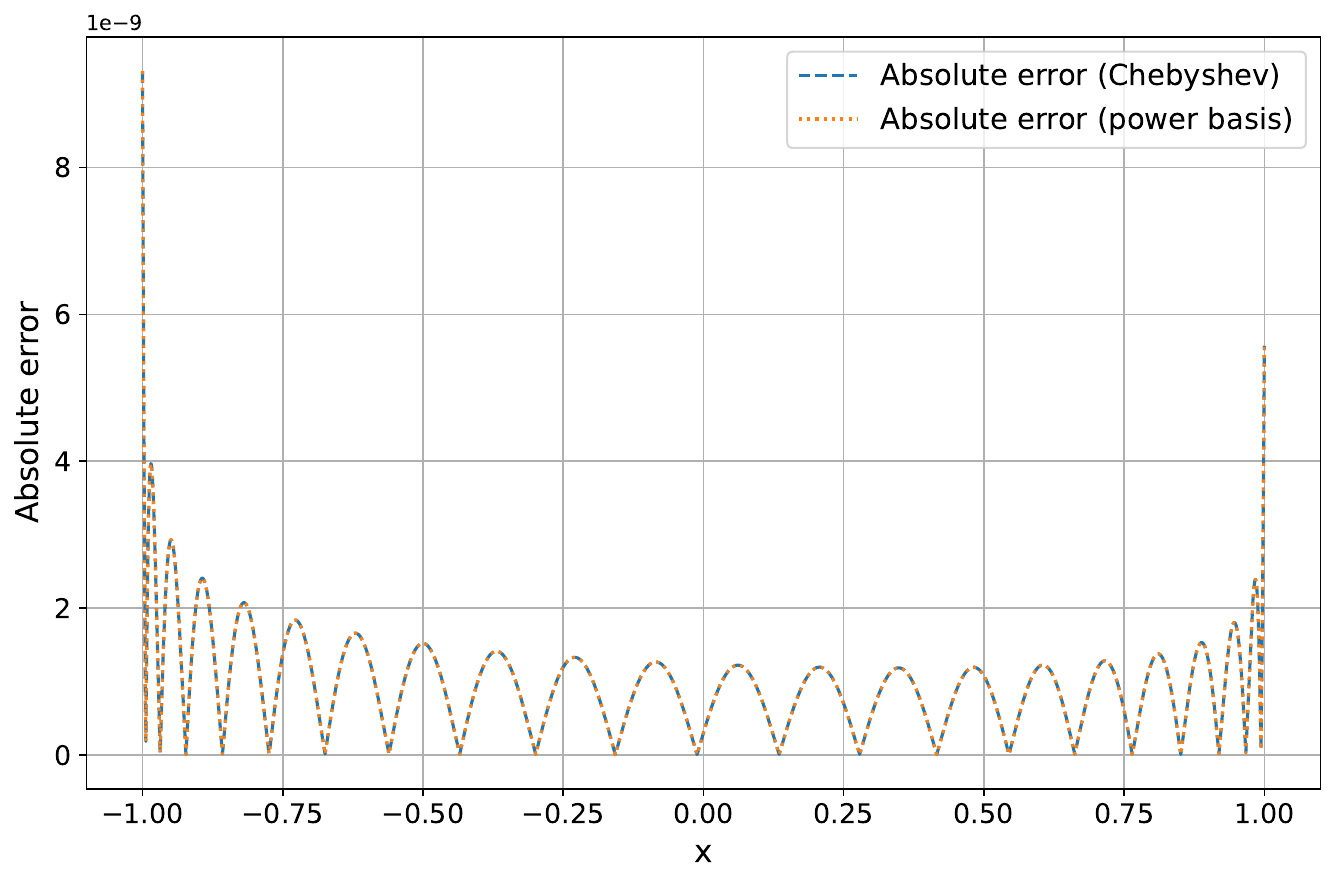}
        \caption{
            Absolute errors of the degree-20 Chebyshev and
            power-basis approximations relative to the exact function
            \(f_{30}(x)\). The nearly coincident error curves demonstrate that
            the two polynomial representations achieve comparable accuracy.
            The errors remain on the order of \(10^{-9}\), with the largest
            deviations occurring near the endpoints of the approximation
            interval.
        }
        \label{fig:finance_poly_error}
    \end{subfigure}

    \caption{
        Accuracy of the degree-20 polynomial approximation of the altered
        inverse function \(f_{30}(x)\). Panel~\textup{(a)} compares the exact
        function with its Chebyshev and power-basis approximations, while
        panel~\textup{(b)} shows their corresponding absolute errors.
    }
    \label{fig:finance_poly_results}
\end{figure}

The specific structure of $M$ allows a cheap reformulation in terms of a Hermitian matrix $H$ using a diagonal similarity transform. In other words, there exists a diagonal matrix $D$ such that
\be
H = D^{-1}\tilde M D.
\ee
The Hermiticity of $H$ reduces to a simple condition on the diagonal entries $d_i$ of D:
\be
\frac{d_{j+1}}{d_j} = \sqrt{\frac{\gamma_{j+1}}{\beta_j}} \label{ratio},
\ee
where we set the zeros in the sub and super diagonals to some small value $\epsilon$ to avoid numerical issues. To enforce $\lVert H \rVert\leq 1$, we scale and shift $H$ as
\begin{equation}
    A = \frac{H-m\mathbb{I}}{h_s},
\end{equation}
with $m = (\lambda_{\max}+\lambda_{\min})/2$ and $h_s =(\lambda_{\max}-\lambda_{\min})/2$. Approximating $H^{-k}$ is equivalent to applying an approximation of the scalar function
\begin{equation}
    f_k(x) = \left(\frac{1}{m+h_sx}\right)^k
\end{equation}
to $A$. As a result, the singularity of this altered inverse is at $x = - \frac{m}{h_s}$. For the positive semi-definite time-step matrix considered here, this lies outside of the interval $[-1,1]$, allowing us to achieve a better approximation, as we never approach the singularity at $x=0$. We now follow the flowchart again, using the Hermitian matrix $A$, and a polynomial that approximates this altered inverse function, shown in \cref{fig:finance_poly_approx}, with errors shown in \cref{fig:finance_poly_error}. 

\begin{figure}[h]
    \centering
    \includegraphics[width=0.8\linewidth]
        {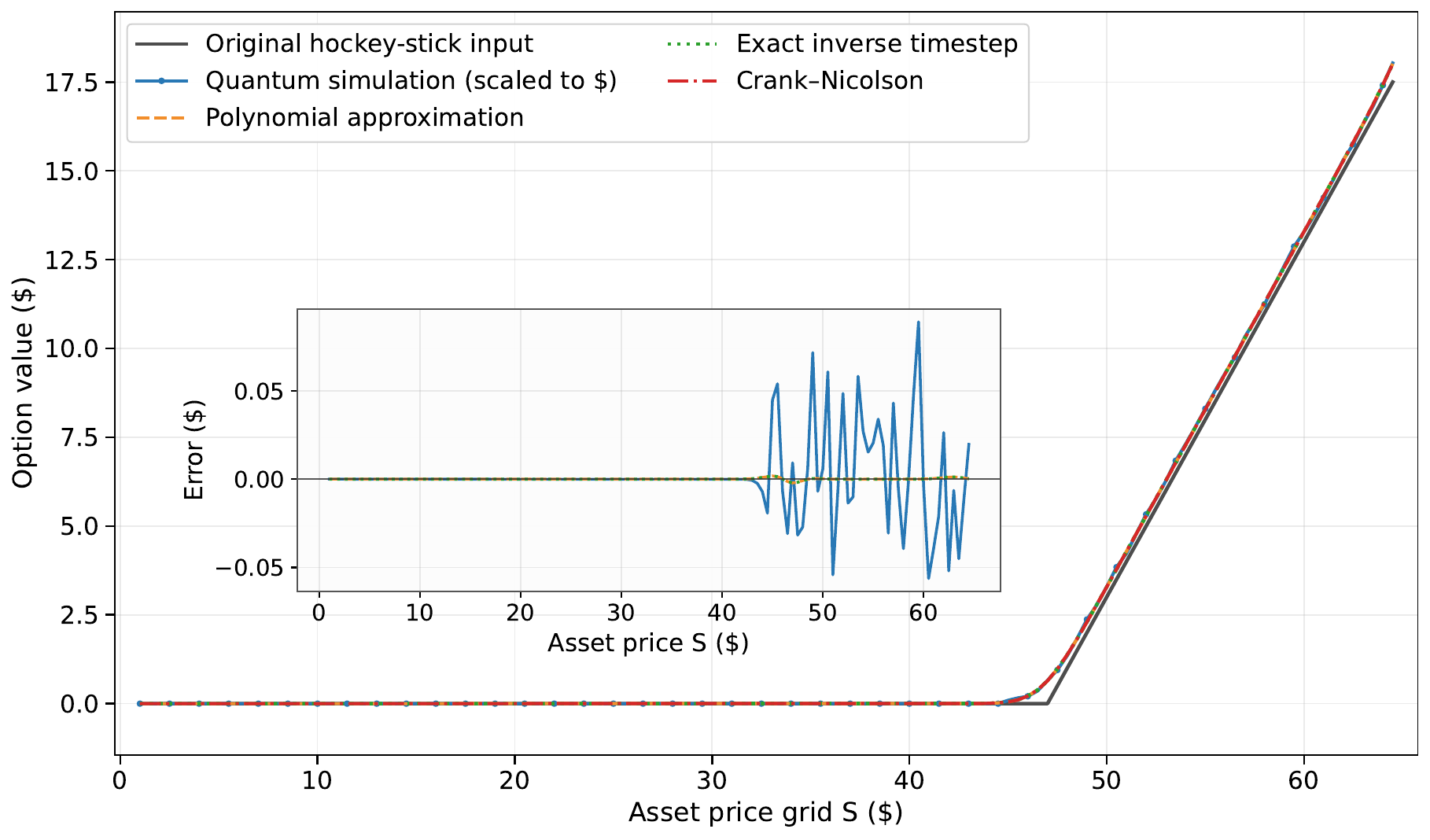}
    \caption{
        Comparison of the original hockey-stick payoff with the quantum,
        polynomial, exact inverse-timestep, and Crank--Nicolson solutions. The
        quantum output is rescaled to dollars. The inset shows the errors
        relative to the Crank--Nicolson solution, with the quantum result
        exhibiting small sampling fluctuations.
    }
    \label{fig:comparison_option_value}
\end{figure}

The obtained option values are presented in \cref{fig:comparison_option_value}, the quantum simulation is compared against a polynomial approximation that inverts the Hermitianized time-step matrix by using the same polynomial approximation as the GQSP, but is applied classically, an ``exact'' inverse was performed by calling Numpy’s linalg.inv() method, applied to the Hermitianized time-step matrix, and then both methods were multiplied with the input vector along with undoing the similarity transformation. The purpose of this is to see how much error is introduced in each step. The polynomial approximation shows how much error GQSP introduces (converting polynomial coefficients to rotation angles), while the exact inverse shows how much error is introduced from the polynomial approximation itself. A crank-Nicolson solution is also provided as comparison to a standard classical solver. The initial success probability of post-selecting the desired block was $\approx 5.76\times10^{-10}$, and after $1000$ calls to oblivious amplitude amplification, the success probability was boosted to $\approx 0.0023$, with the optimal number of calls being 32730.

\vspace{1cm}
\begin{tcolorbox}[colback=blue!5!white,colframe=blue!75!black,title=Flowchart decisions for the Black--Scholes equation example]
\begin{enumerate}
    \item Is the matrix square? \\ Yes, the time-step matrix $\tilde{M}$ is square.
    \item Is the matrix unitary? \\No, generally the matrix $\tilde{M}$ will not be unitary.
    \item Is the matrix normal? \\No, generally the matrix $\tilde{M}$ will not be normal.
    \item Does the function need to be applied to the singular values of $\tilde{M}$? \\No, we do not wish to apply the function $f$ to the singular values of $\tilde{M}$.
    \item Thus we must reconsider, a path forward in this case is to Hermitianize $\tilde{M}$ such that we obtain a Hermitian matrix $A$, and start from the top of the flowchart again.
    \item Is the matrix square? \\ Yes, the Hermitianized time-step matrix $A$ is square.
    \item Is the matrix unitary? \\No, generally the matrix $A$ will not be unitary.
    \item Is the matrix normal? \\Yes, the Hermitianized time-step matrix $A$ is normal.
    \item Can we decompose the matrix into a Laurent polynomial? \\ Yes, since $A$ is Hermitian with $\lVert A \rVert \leq 1$, it can be decomposed as $A = \dfrac{1}{2}U + \dfrac{1}{2}U^\dagger$ with $U = A + i\sqrt{I - A^2}$. Note that this $U$ is not necessarily efficiently implementable in a quantum circuit, so this route must be compared with the alternative option: qubitization. For this simulated example, we do not take this efficiency into account.
    \item Is the GQSP condition satisfied? \\ Yes, however $P$ may optionally be normalized so that its maximum approaches one, thereby increasing the post-selection success probability.
    \begin{tcolorbox}[colback=yellow!5!white,colframe=yellow!75!black,title=GQSP Condition]
    \[
        |P(e^{i \theta})|\leq 1,\ \forall \theta \in[0,2\pi].
    \]
    \end{tcolorbox}
    \item Apply GQSP.
\end{enumerate}
\end{tcolorbox}

\addtocontents{toc}{\string\setcounter{tocdepth}{1}}

\newpage
\section{Open Problems, Future Directions and Conclusion}

\subsection{Open Problems and Future Directions}

Despite the remarkable progress enabled by block-encoding, qubitization, quantum signal processing (QSP), quantum singular value transformation (QSVT), and generalized quantum signal processing (GQSP), quantum linear algebra remains a rapidly evolving field with numerous open challenges. These challenges span mathematical foundations, algorithmic design, computational complexity, resource estimation, and practical implementation. Many of the most promising future directions arise from attempts to reduce overheads, broaden applicability, and establish practical quantum advantages for scientifically relevant problems.

The most pressing open problem remains whether GQSP can be extended into the singular value domain, with arbitrary matrices, in the same way as QSP was extended into QSVT. Since the two frameworks work under fundamentally different constraints, a generalized quantum singular value transform which removes the parity and real-coefficient constraints would be the gold standard signal processing approach. However, it is not known whether an efficient extension which resolves these constraints is possible.

Block-encoding is the cornerstone of modern quantum linear algebra. However, constructing efficient block-encodings often constitutes the dominant cost of an algorithm. Many theoretical quantum speedups assume the availability of efficient block-encodings, making their practical realization one of the most important outstanding challenges. While there are now several resources \cite{camps2024blockencodings,li2023efficient,nibbi2024blockencoding,sunderhauf2024blockencoding,wan2021blockencoding,lapworth2025precondition} which outline block-encodings for different matrices, generally block-encodings are problem-specific in order to exploit a structure present in the matrix. This presents a challenge in developing efficient general block-encodings.

For QSP and QSVT, phase rotation angle synthesis had been a large classical bottleneck, especially for polynomials of very large degree. Although now the phase angle synthesis algorithm described in \cite{motlagh2024generalized} has alleviated the problem for GQSP, QSP and QSVT, the problem has shifted to efficiently finding a complementary polynomial. The complementary polynomial can be found by various methods including numerical optimization \cite{motlagh2024generalized}, root finding \cite{bergholm2022pennylane}, factorization \cite{ying2022stable}, or novel applications of the FFT \cite{Berntson2025complementary}. It remains an open area of research to improve the efficiency of complementary polynomial finding algorithms.

\subsection{Conclusion}

Just as classical linear algebra underpins modern scientific computing, data science, and machine learning, quantum linear algebra is emerging as the mathematical foundation of next-generation quantum algorithms. By exploiting quantum superposition, interference, and entanglement, quantum computers can represent and manipulate high-dimensional vectors and operators in ways that are fundamentally different from classical computation. Consequently, many of the most powerful quantum algorithms can be interpreted as sophisticated linear-algebraic procedures acting on quantum states, with tasks such as matrix inversion, eigenvalue estimation, singular value transformation, and operator simulation lying at the heart of quantum computational advantage.

Quantum linear algebra has therefore become one of the central pillars of modern quantum algorithm development. A wide range of quantum algorithms can be viewed as performing transformations on vectors, matrices, and linear operators encoded within quantum states. Over the past two decades, the field has evolved from early breakthroughs such as the Harrow–Hassidim–Lloyd (HHL) linear systems algorithm and quantum walk techniques into a mature framework built upon Hamiltonian simulation, block-encoding, qubitization, quantum signal processing (QSP), quantum singular value transformation (QSVT), and generalized quantum signal processing (GQSP). Together, these methods provide a powerful and versatile toolkit for implementing matrix functions, solving linear systems, estimating eigenvalues and singular values, simulating dynamical systems, and performing polynomial transformations of operators with provably optimal or near-optimal resource requirements.

A key strength of quantum linear algebra is its ability to unify seemingly diverse quantum algorithms within a common mathematical framework. Problems arising in optimization, machine learning, network science, computational finance, scientific computing, differential equations, quantum chemistry, and many-body physics can often be reformulated as questions involving matrix functions or operator transformations. Rather than designing bespoke quantum algorithms for each application, researchers can increasingly leverage a common set of linear-algebraic primitives that can be systematically combined and adapted to new domains. This shift has transformed quantum algorithm design from a collection of specialized techniques into a more coherent and principled discipline.

Within this framework, block-encoding, qubitization, QSP, QSVT, and GQSP form a natural hierarchy of algorithmic abstractions. Block-encoding provides a mechanism for representing arbitrary matrices within larger unitary operators, thereby making non-unitary linear-algebraic objects accessible to quantum computation. Qubitization then enables efficient access to the spectral properties of these encoded operators, converting matrix manipulations into controlled quantum-walk dynamics. Quantum Signal Processing offers an optimal method for implementing polynomial transformations of scalar eigenvalues using carefully engineered phase sequences. Quantum Singular Value Transformation generalizes this capability to arbitrary matrices, enabling polynomial transformations of singular values and thereby unifying numerous quantum algorithms under a single theoretical framework. More recently, Generalized Quantum Signal Processing has expanded the range of achievable operator transformations and practical constructions, offering additional flexibility for implementing matrix functions beyond the standard QSP and QSVT paradigms.

Collectively, these developments have reshaped the landscape of quantum algorithm research. Many problems that were previously approached through ad hoc algorithmic constructions can now be formulated as instances of operator representation, spectral access, and polynomial transformation. In this sense, quantum linear algebra provides not only a collection of algorithms but also a general methodology for quantum algorithm design. The resulting framework has already influenced fields as diverse as Hamiltonian simulation, quantum chemistry, combinatorial optimization, machine learning, quantum finance, and scientific computing, and it is expected to play an even greater role as fault-tolerant quantum hardware becomes available.

Looking forward, quantum linear algebra is likely to remain a cornerstone of large-scale quantum computing. Significant challenges remain, including the efficient construction of block-encodings, the synthesis of numerically stable phase sequences, realistic fault-tolerant resource estimation, and the identification of application domains where asymptotic quantum speedups translate into practical computational advantages. Addressing these challenges will be crucial for bridging the gap between theoretical algorithmic advances and real-world quantum applications. Nevertheless, the rapid development of quantum linear algebra over the past decade strongly suggests that it will continue to serve as one of the most important conceptual and technical foundations for future quantum algorithm research.

\subsection{Acknowledgment}
This research is supported by the Critical Technologies Challenge Program of the Australian
Government Department of Industry, Science and Resources. Numerical simulations
were conducted on Setonix, provided by the Pawsey Supercomputing Research Centre. KDS, AM, JR are supported 
by the European Union (UE) and Region Reunion (FEDER).

\bibliography{bibliography}

\end{document}